\documentclass[acmsmall]{acm/acmart}
\AtBeginDocument{%
  }

\setcopyright{acmlicensed}
\copyrightyear{2026}
\acmYear{2026}
\acmDOI{XXXXXXX.XXXXXXX}

\acmJournal{JACM}
\acmVolume{1}
\acmNumber{1}
\acmArticle{1}
\acmMonth{3}

\usepackage[breakable]{tcolorbox}
\usepackage{booktabs}  % 用于三线表 (\toprule, \midrule, \bottomrule)
\usepackage{tabularx}  % 用于自动换行的列 (X column)
\usepackage[table]{xcolor} % 用于表格行颜色
\usepackage{multirow} % For multirow cells in tables
\usepackage{threeparttable}
\usepackage{adjustbox}
\usepackage{xparse}
\usepackage{graphicx}
\usepackage{spverbatim}
\usepackage{xcolor}
\usepackage{makecell}
\usepackage{enumitem}
\usepackage{appendix}
\usepackage{listings}
\newcommand{\lstbg}[3][0pt]{{\fboxsep#1\colorbox{#2}{\strut #3}}}

\lstdefinelanguage{diff}{
  numbers=left,
  numberstyle=\tiny\color{gray},
  stepnumber=1,
  numbersep=5pt,
  frame=single,
  backgroundcolor=\color{gray!5},
  basicstyle=\ttfamily\scriptsize,
  morecomment=[f][\lstbg{red!12}]-,
  morecomment=[f][\lstbg{green!12}]+,
  morecomment=[f][\textit]{@@},
}

\newcommand{\tech}{RefUntangle}

\NewDocumentCommand{\framecolorbox}{oommm}
 {% #1 = width (optional)
  \IfValueTF{#1}
   {\IfValueTF{#2}
    {\fcolorbox{#3}{#4}{\makebox[#1][#2]{#5}}}
    {\fcolorbox{#3}{#4}{\makebox[#1]{#5}}}%
   }
   {\fcolorbox{#3}{#4}{#5}}%
 }

\newif\ifincludeappendix
\includeappendixtrue

\begin{document}

%%
%% The "title" command has an optional parameter,
%% allowing the author to define a "short title" to be used in page headers.
% \title{An Empirical Study of coding agent Refactoring in Tangled Commits on Multi-SWE-bench Java Projects}
% \title{An Empirical Study of Tangled Refactorings in LLM-based Agents for Software Issue Resolution}
\title{``Refactoring Runaway'': Understanding and Mitigating Tangled Refactorings in Coding Agents for Issue Resolution}

%%
%% The "author" command and its associated commands are used to define
%% the authors and their affiliations.
%% Of note is the shared affiliation of the first two authors, and the
%% "authornote" and "authornotemark" commands
%% used to denote shared contribution to the research.
\author{Zhao Tian}
\authornote{Both authors contributed equally to this research.}
\orcid{0000-0002-9316-7250}
\affiliation{
  \institution{School of Computer Software, Tianjin University}
  \city{Tianjin}
  \country{China}
}
\email{tianzhao@tju.edu.cn}

\author{Zifan Zhang}
\authornotemark[1]
\email{template@}
\affiliation{
  \institution{Kyushu University}
  \city{Fukuoka}
  \country{Japan}
}
\email{zifan.614@s.kyushu-u.ac.jp}

\author{Tao Xiao}
\authornote{Tao Xiao is the corresponding author.}
\orcid{0000-0003-4070-585X}
\affiliation{
  \institution{Kyushu University}
  \city{Fukuoka}
  \country{Japan}
}
\email{xiao@ait.kyushu-u.ac.jp}

\author{Dong Wang}
\orcid{}
\affiliation{
  \institution{School of Computer Software, Tianjin University}
  \city{Tianjin}
  \country{China}
}
\email{dong_w@tju.edu.cn}

\author{Masanari Kondo}
\orcid{}
\affiliation{
  \institution{Kyushu University}
  \city{Fukuoka}
  \country{Japan}
}
\email{kondo@ait.kyushu-u.ac.jp}

\author{Junjie Chen}
\orcid{0000-0003-3056-9962}
\affiliation{
  \institution{School of Computer Software, Tianjin University}
  \city{Tianjin}
  \country{China}
}
\email{junjiechen@tju.edu.cn}

\author{Yasutaka Kamei}
\orcid{0000-0002-7058-1045}
\affiliation{%
  \institution{Kyushu University}
  \city{Fukuoka}
  \country{Japan}
}
\email{kamei@ait.kyushu-u.ac.jp}

%%
%% By default, the full list of authors will be used in the page
%% headers. Often, this list is too long, and will overlap
%% other information printed in the page headers. This command allows
%% the author to define a more concise list
%% of authors' names for this purpose.
% \renewcommand{\shortauthors}{Trovato et al.}

%%
%% The abstract is a short summary of the work to be presented in the
%% article.
\begin{abstract}
Recent advances in coding agents have shown remarkable progress in software issue resolution. 
In practice, real-world issues are typically bug fixes or feature requests in which human developers naturally incorporate refactoring as part of the resolution process, resulting in tangled refactoring. 
Since LLMs are trained on large-scale open-source repositories, coding agents may inherit such behaviors.
In this paper, we conduct an empirical study on Multi-SWE-bench, analyzing 3,691 valid patches generated by three agent frameworks with 12 LLMs. 
We find that coding agents introduce tangled refactorings less frequently (21.43\% \textit{vs.} 36.72\%) and with lower intensity (0.66 \textit{vs.} 1.75) than human developers, although they exhibit a broader diversity of refactoring types. 
Logistic regression analysis further shows that tangled refactorings are strongly associated with reduced compilability, while exhibiting no significant association with functional correctness. 
Based on these findings, we propose a refactoring-aware refinement approach that assesses the necessity and safety of tangled refactorings and selectively removes or repairs problematic operations. 
Our approach improves compilability from 19.34\% to 38.33\%, and additionally resolves 2.79\% previously unresolved issues. 
Overall, this work presents the first step towards understanding tangled refactoring practices in agentic issue resolution and opens up avenues for future work.

\end{abstract}

%%
%% The code below is generated by the tool at http://dl.acm.org/ccs.cfm.
%% Please copy and paste the code instead of the example below.
%%

% \ccsdesc[500]{Software and its engineering}

%%
%% Keywords. The author(s) should pick words that accurately describe
%% the work being presented. Separate the keywords with commas.

% \received{20 February 2007}
% \received[revised]{12 March 2009}
% \received[accepted]{5 June 2009}

%%
%% This command processes the author and affiliation and title
%% information and builds the first part of the formatted document.
\maketitle

\section{Introduction}
Software issue resolution is the automated task of understanding real-world software issues, locating their underlying causes, and generating patches to resolve them.
Such issues typically arise in the form of bug reports or feature requests, and their resolution constitutes a fundamental aspect of software maintenance~\cite{bijlsma2012faster,li2024bug,wang2025survey, gao2025trae,chen2026icse} due to their practical impact on the software development lifecycle. 
Recently, SWE-bench~\cite{jimenez2024swebench}, which comprises 2,294 issue instances collected from 12 well-maintained Python repositories, has emerged as a prominent benchmark for evaluating this task~\cite{martinez2026s}. 
Beyond providing a large-scale and realistic assessment framework~\cite{jin2024llms}, SWE-bench has also accelerated the evolution of Large Language Models (LLMs) from code completion assistants to increasingly autonomous software engineering agents~\cite{jiang2025agentic,martinez2026s}, as exemplified by systems such as SWE-agent~\cite{sweagent} and OpenHands~\cite{wang2025openhands}.

% Transited from SWE-bench to its evaluation and patch analysis. Most SOTA evaluation and empirical studies focused on the functional correctness of the generated patches, overlooking the non-functional code quality of the generated patches. Mentions one work for the issue resolution task, agents increase complexity.
While state-of-the-art LLMs and coding agents achieve impressively high issue resolution rates (measured by their ability to generate patches that pass pre-defined test suites) on SWE-bench and its variants~\cite{martinez2026s}, existing evaluations and analyses predominantly focus on functional correctness. 
Consequently, the non-functional characteristics of agent-generated patches have only recently begun to receive research attention~\cite{sajadi2025safe,chen2025evaluating,sun2025quality}. 
For example, \citet{sajadi2025safe} found that coding agents with greater autonomy can further amplify the risk of introducing vulnerability.
\citet{chen2025evaluating} observed that while coding agents generally preserve code reliability and security during issue resolution, they frequently introduce additional structural complexity into the codebase.
% \wang{add another example here to enrich.}

% Transited from agents for issue resolution increases complexity to traditional software developers, introducing refactoring, tangled commits, and tangled refactoring is one particular entangled chnages
In typical software development, developers often manage structural complexity through refactoring, which restructures code without changing its external behavior~\cite{fowler2018refactoring}.
Although software engineering best practices recommend that refactorings should be submitted as atomic, single-purpose commits, real-world development frequently deviates from this principle.
Developers commonly produce tangled commits that combine multiple change intents within a single issue resolution task, such as mixing bug fixes with structural modifications~\cite{herzig2013impact,tao2015untangling,Bacchelli2015untangling}.
Among these mixed changes, \textbf{tangled refactoring}~\cite{niu2025refactoring}, where refactorings are interleaved with other change types, is particularly prevalent~\cite{floss_refactoring_definition,kirinuki2014tangled,Bacchelli2015untangling}.
This phenomenon naturally arises because developers often perform opportunistic structural cleanups while addressing a primary maintenance task~\cite{Herboldfinegrained}.

\begin{figure}[t]
    \centering
    \includegraphics[width=\linewidth]{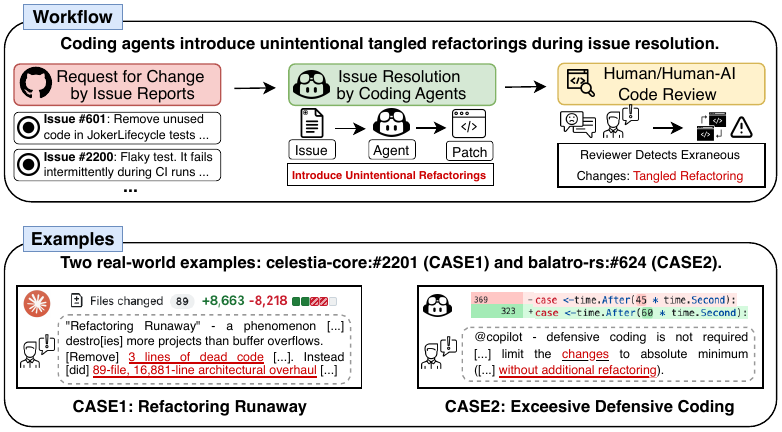}
    \caption{Real-world examples of unintentional refactorings performed by coding agents during issue resolutions.}
    \label{fig:mot}
\end{figure}

% Transited from tangled refactoring is human development, propose a hypothesis whether coding agents for issue resolution learn untangled refactoring from humans because of their training data. Coin a new term for our context, unintentional refactoring. Using a motivating example to illustrate the two unintentional refactoring in the real world that were made by agents for issue resolution, pointing out its nesscity to conduct this study. 
Given that coding agents are powered by LLMs trained on massive open-source software repositories, where tangled refactorings are widespread, they may inherit not only common development practices, but also suboptimal maintenance behaviors embedded in real-world commits.
During automated issue resolution, this tendency can manifest as \textit{unintentional refactoring}: structural modifications introduced by agents that are neither explicitly requested nor implicitly required by the issue itself.
Figure~\ref{fig:mot} presents representative examples of this phenomenon, where coding agents perform excessive and unsolicited structural edits during otherwise straightforward issue resolution tasks.
In \textit{CASE1},\footnote{\url{https://github.com/spencerduncan/balatro-rs/pull/624\#issuecomment-3127094454}} the issue merely requests the removal of three lines of dead code.
Instead, the agent triggers a \emph{Refactoring Runaway}, generating 16,881 lines of structural modifications across 89 files.
The resulting changes are so disruptive that developers compare their impact to a critical vulnerability ``buffer overflow,'' underscoring the potential danger of uncontrolled AI-driven refactoring behavior.
In \textit{CASE2},\footnote{\url{https://github.com/celestiaorg/celestia-core/pull/2201\#issuecomment-3113680978}} the task simply requires increasing a timeout value to repair a flaky test.
The excessive modifications significantly burden the review process, ultimately forcing developers to intervene and explicitly constrain the agent's behavior: \textit{``defensive coding is not required in test code. Please limit the changes to absolute minimum.''}
Developers further demand the removal of the bloated refactoring code before the patch could proceed.
These examples reveal a critical misalignment between the minimal intent expressed in issue reports and the actual repair behavior exhibited by coding agents.
Rather than performing focused issue resolution, agents may expand simple maintenance tasks into large-scale, tangled refactoring activities that increase review cost, reduce change interpretability, and potentially introduce unintended risks.
Despite these implications, unintentional tangled refactorings introduced by coding agents remain unexplored.

% To the best of our knowledge, no prior studies have systematically investigated whether \textbf{LLM-powered coding agents mimic human developers by producing unintentional tangled refactorings during issue resolution}. 
% Understanding this phenomenon is necessary because unprompted refactoring by autonomous agents may lead to scope creep, complicate human code review, and increase the risk of breaking Continuous Integration (CI) pipelines through tangled refactorings. 
To address this gap, this paper presents the first empirical study of unintentional tangled refactorings generated by coding agents during automated issue resolution to systematically characterize this phenomenon and explore its potential risks and mitigation opportunities.
Using the Multi-SWE-bench dataset~\cite{zan2025multiswebench}, we investigate the prevalence of refactoring-related tangling, analyze the structural characteristics of the generated refactorings, evaluate their impact on patch correctness, and explore a mitigation strategy.
To guide this study, we firstly formulate the following two research questions (RQs):

\begin{itemize}
    \item \textbf{RQ1: How do tangled refactorings in issue-resolution patches generated by coding agents differ from those by human developers?} 
    Previous studies on human software development have shown that tangled refactoring is a pervasive phenomenon, with approximately 15.80\% of commits in just-in-time defect prediction datasets~\cite{niu2025refactoring}.
    However, it remains unclear whether coding agents exhibit similar tangling behaviors during automated issue resolution and whether their refactoring patterns differ from those of human developers.
    Therefore, RQ1 investigates the prevalence and characteristics of tangled refactorings in agent-generated patches, establishing a comparative baseline against developer behavior.

    \item \textbf{RQ2: How do tangled refactorings affect the effectiveness of coding agents in issue resolution?} 
    Unintentional structural edits introduced during issue resolution may increase patch complexity and divert coding agents from the primary repair objective. 
    In practice, excessive or unnecessary refactorings could make generated patches harder to validate, review, and compile correctly.
    Therefore, RQ2 investigates the relationship between tangled refactorings and the effectiveness of agent-generated patches by statistically analyzing their association with code compilability and issue resolution success.
\end{itemize}

Based on our empirical analysis, we derive the following key findings.
For \textbf{RQ1}, our results show that human developers perform tangled refactorings more frequently and intensively than coding agents during issue resolution. 
Specifically, 36.72\% of human-written patches contain tangled refactorings, with an average of 1.75 tangled refactorings per patch, whereas only 21.43\% of agent-generated patches contain tangled refactorings, averaging 0.66 per patch.
However, coding agents exhibit substantially greater diversity in tangled refactorings, covering 73 distinct tangled refactoring types compared to 46 for human developers.
In particular, coding agents tend to perform fine-grained expression-level transformations such as \textit{Extract Variable}, while human developers more commonly apply method-level structural refactorings such as \textit{Extract Method}.
For \textbf{RQ2}, our logistic regression analysis shows that tangled refactorings significantly reduce patch compilability, but do not statistically significantly improve the likelihood of successful issue resolution.
In particular, method-level tangled refactorings are negatively associated with agent effectiveness, reducing both issue resolution and compilation success, whereas removal-oriented tangled refactorings exhibit the opposite effect.

Motivated by these findings and recent advances in automated patch refinement~\cite{pabba2025refineenhancingprogramrepair}, we further explore whether automated techniques can mitigate these refactoring-induced errors, such as syntactic or dependency-level instability in agent-generated patches. 
Accordingly, we formulate the following RQ:

\begin{itemize}
    \item \textbf{RQ3: How effective is \tech{} in improving coding agents for issue resolution?} 
    To address the negative effects of tangled refactorings identified in RQ2, we propose \tech{}, a refactoring-aware refinement approach that first assesses the quality of tangled refactoring operations in agent-generated patches, and then selectively modifies or removes low-quality, unsafe, or unnecessary refactorings.
    RQ3 evaluates whether \tech{} can mitigate problematic tangled refactorings and thereby improve the compilability and functional correctness of agent-generated patches.
\end{itemize}

The LLM-based assessment reveals that 50.44--71.28\% of agent-generated patches contain at least one unnecessary or unsafe refactoring, highlighting the widespread presence of refactoring-related issues in current coding agent's outputs.
Our results further show that \tech{} effectively mitigates these structural instabilities, nearly doubling patch compilability by increasing the average compilation success rate from 19.34\% to 38.33\%.
Moreover, \tech{} enables an additional 2.79\% of previously unresolved patches to pass all tests, suggesting that refactoring-aware structural refinement can also improve functional correctness of patches.

\smallskip 
\noindent
% By answering these research questions, this study provides empirical insights into how coding agents implicitly modify code structure in realistic software maintenance scenarios, how such behavior affects patch correctness, and how it can inform the design of future agentic frameworks\tao{, such as (1) restricting large-scale refactorings unless explicitly required by the issue context; (2) validating method-signature modifications against inheritance hierarchies and interface contracts; and (3) separating refactoring-related transformations from bug-fix generation into independently verifiable stages}\wang{elaborate on a bit further, what concrete suggestions can be provided?}.
% modify code with implicit or unintentional refactoring in realistic scenarios and how their tangled refactoring behavior relates to patch correctness in issue resolution tasks and future improvements in agentic framework design.
\textbf{Contributions.} 
This paper makes the following key contributions: 
(1) We present the first empirical study of tangled refactorings in coding agent-generated patches, systematically characterizing their prevalence, structural patterns, and impact on issue resolution.
(2) We propose and evaluate \tech{}, a refactoring-aware two-stage patch refinement pipeline that mitigates problematic tangled refactorings and improves patch compilability and correctness.
(3) Based on our empirical findings, we derive actionable insights for both researchers and practitioners. 
To facilitate future research and reproducibility, we release a public replication package~\cite{replication}.
% \footnote{\url{https://github.com/zifan-zhang/llm-refactoring-research}}

% , based on 3,691 valid patches from three agent frameworks pairs with 12 different LLMs in Multi-SWE-bench~\cite{zan2025multiswebench}. (2) We establish a clear empirical link between tangled refactoring and patch correctness: the presence of tangled refactoring is strongly associated with lower code compilability, exhibits no significant overall association with issue resolution success, and demonstrates varying effects across different refactoring dimensions (action and scope). (3) We propose and validate \tech, a refactoring-aware two-stage patch refinement pipeline. This approach nearly doubles patch compilability (from 19.34\% to 38.33\%), enables an additional 2.79\% of previously unresolved patches to pass all tests, and reveals that 50.44--71.28\%  of initial agent-generated patches contain unnecessary or unsafe refactorings. To facilitate reproducibility and support future research, we publicly release a comprehensive replication package, available at \textcolor{violet}{\url{https://github.com/zifan-zhang/llm-refactoring-research}}.

\section{Related Work}
\subsection{Tangled Refactoring}

Prior research highlights that refactoring is rarely performed in isolation. 
\citet{floss_refactoring_definition} distinguish between \emph{root-canal} refactoring, performed in isolation, and \emph{floss} refactoring, which is interleaved with functional changes. 
This interleaving leads to the phenomenon of ``tangled refactoring,'' where refactoring is mixed not only with functional changes but also with bug fixes or the implementations of new functionality.
% \citet{herzig2016impact} and \citet{herzig2013impact} demonstrate that up to 15\% of bug-fixing commits contain tangled changes, introducing significant noise into empirical analyses. 
% Further granularity is provided by 
\citet{Herboldfinegrained} reported that only about half of the code within bug-fixing commits directly contributes to the fix, with the remainder often attributed to refactorings and unrelated modifications. 
Additionally, \citet{negara2013comparative} find that many such refactorings are performed manually rather than through automated IDE tools, increasing the likelihood of human error during these interleaved tasks.

While refactoring aims to improve internal structure, it often introduces unintended side effects across various dimensions of software development. The relationship between refactoring and bug introduction has been extensively studied. 
\citet{Bavota_bug_refactoring} and \citet{Counsell_bug_refactoring} identify that specific refactoring types, particularly hierarchy-related operations, are frequently associated with the introduction of new defects. 
\citet{dipentarefactoringbugs} observe a temporal link where refactoring actions are often followed by subsequent defect corrections. 
Moreover, \citet{Bagheri2022RefactorBugs} find that a large fraction of bug-inducing commits co-occur with refactoring operations, suggesting that refactoring is not always behavior-preserving in practice. Beyond defects, refactoring complicates collaborative development and quality assurance. 
\citet{Mahmoudi_refactoring_merge_conflicts} report that refactoring operations are involved in 22\% of merge conflicts, and these conflicts tend to be more complex than those without refactoring. 
Refactoring also poses risks to the stability of the test suite; for instance, \citet{doesrefactoringbreaktests} find that structural changes like renaming attributes or classes have a high probability of breaking existing tests. 

Our work extends this line of research by applying refactoring detection to AI-generated patches rather than human-written commits. By comparing the refactoring practices of coding agents with those of human developers, we offer new insights into how LLM-based code generation interacts with established software engineering patterns.

\subsection{Automated Issue Resolution by Coding Agents}
Large Language Models (LLMs) have enabled autonomous coding agents that can resolve real-world software engineering issues.
\citet{jimenez2024swebench} established SWE-bench, a widely adopted evaluation framework using GitHub issues from Python repositories, later extended to multiple languages by \citet{zan2025multiswebench} with Multi-SWE-bench.

Several agent frameworks have been proposed to tackle these benchmarks.
\citet{sweagent} grant LLMs high autonomy through a custom shell environment for file navigation and iterative editing in SWE-agent.
\citet{wang2025openhands} adopt a similar interactive sandboxed design with multi-turn interactions in OpenHands.
\citet{agentless} follow a structured three-phase pipeline (localization, repair, validation) in Agentless, constraining autonomy in favor of reproducibility.
Other notable systems include AutoCodeRover by \citet{autocoderover}, which combines code search with program analysis, and MAGIS by \citet{MAGIS}, which orchestrates multiple specialized agents.

\citet{ceka2025understandingsoftwareengineeringagents} provide an empirical comparison of how these agents explore and modify codebases, finding that behavior varies substantially across frameworks.
Concurrent with our work, recent studies have begun to examine the refactoring behavior of AI coding agents directly.
\citet{horikawa2025agenticrefactoringempiricalstudy} conduct a large-scale analysis of 15,451 intentional refactoring instances across 12,256 pull requests in open-source Java projects, finding that agent refactoring is dominated by low-level, consistency-oriented edits such as \textit{Change Variable Type} and \textit{Rename Parameter}, in contrast to the diverse structural improvements typical of human developers.
Similarly, \citet{ottenhof2026howagentsrefactor} compare agent-generated and developer-driven refactoring pull requests across 86 Java projects, reporting that agent refactorings are heavily concentrated on annotation-related changes.
However, both studies focus on \emph{intentional} refactoring where agents are explicitly tasked with code restructuring.
In contrast, our work examines \emph{implicit} refactoring that is tangled with issue resolution patches and investigates its characteristics, and impact on functional correctness.

While significant effort has been devoted to improving the issue resolution rate, relatively little attention has been paid to the \emph{structural quality} of generated patches.
Prior evaluations, including those by \citet{agentless} and \citet{MAGIS}, measure functional correctness (i.e., whether the patch passes the test suite) without examining how agents modify the codebase beyond the minimal fix.
Our study fills this gap by systematically analyzing the refactoring operations embedded within agent-generated patches for real-world issue resolution task and their impact on patch correctness.

\section{Dataset Preparation}
\label{sec:dp}

\begin{figure}[t]
    \centering
    \includegraphics[width=1\linewidth]{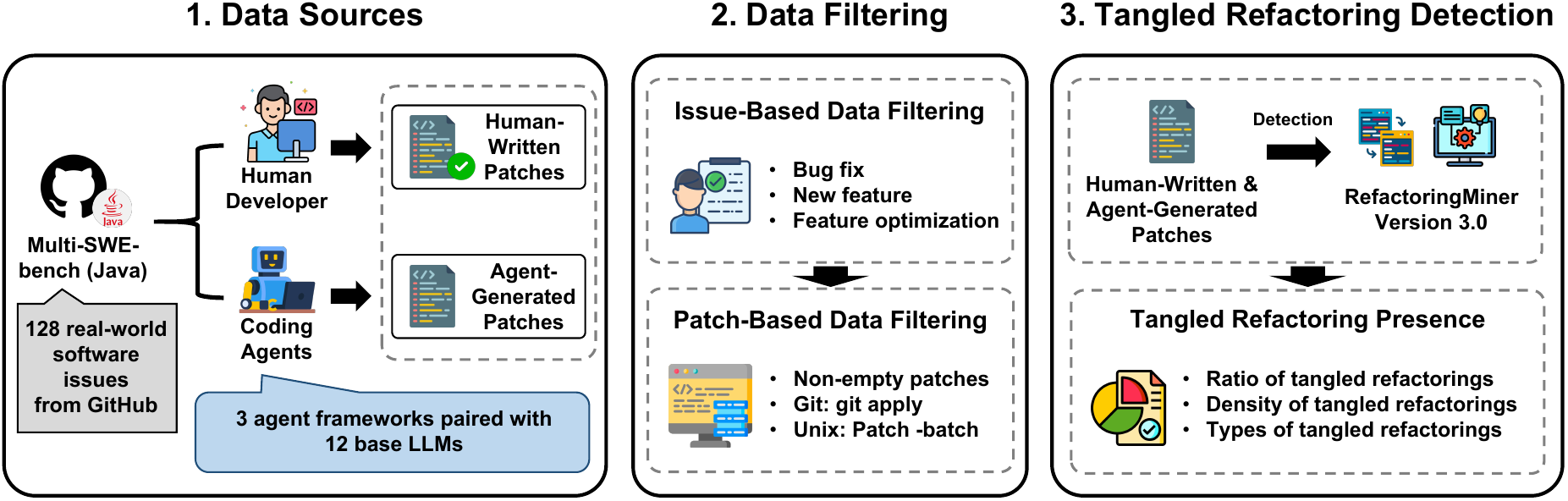}
    \caption{The dataset preparation process for empirical analysis of tangled refactorings in coding agents for issue resolution.}
    \label{fig:overall_design}
\end{figure}
% Figure~\ref{fig:overall_design} illustrates the overall design of our study. 
In this section, we introduce the dataset preparation process for the empirical analysis of tangled refactorings in coding agents for issue resolution, illustrated in Figure~\ref{fig:overall_design}.
% Using both human-written and agent-generated patches from the widely-used software issue resolution benchmark (i.e., Multi-SWE-bench~\cite{zan2025multiswebench}), we first conduct a comparative analysis of refactoring practices between these two types of patches (\textbf{RQ1}). 
% Next, we employ statistical modeling to investigate how tangled refactorings influence the compilability and correctness of patches in the context of issue resolution (\textbf{RQ2}).
% Finally, we design a two-stage pipeline to assess and refine agent-generated patches, with the goal of improving their compilability and correctness, thereby enhancing overall issue resolution performance (\textbf{RQ3}).

\subsection{Data Sources}
To analyze refactoring operations performed by coding agents in real-world software maintenance scenarios for issue resolution, we adopt the widely-used benchmark Multi-SWE-bench~\cite{zan2025multiswebench}. 
This benchmark extends the well-known SWE-bench~\cite{jimenez2024swebench} and is designed to evaluate coding agents on their ability to resolve real-world software issues.
In this study, we focus specifically on the Java programming language, which includes 128 issue instances in Multi-SWE-bench. 
This choice enables us to leverage the state-of-the-art refactoring detection tool RefactoringMiner~\cite{Alikhanifard:TOSEM:2024:RefactoringMiner3.0}, which currently supports only Java source code.
Recently, \citet{yang2025swesmith} introduced SWE-bench Multilingual, a multi-programming-language extension of SWE-bench that also includes Java instances. 
However, it contains only 43 Java issues. 
Therefore, we select Multi-SWE-bench, which provides a larger and more diverse set of relevant Java instances to support our comprehensive analysis.

The Multi-SWE-bench benchmark contains 128 human-written golden patches, which serve as the ground truth for correct issue resolution.
In addition, its public repository provides 4,608 agent-generated patches.
These 4,608 patches (3 agent frameworks $\times$ 12 LLMs $\times$ 128 software issues) were produced by three state-of-the-art coding agents (i.e., SWE-agent~\cite{sweagent}, OpenHands~\cite{wang2025openhands}, and Agentless~\cite{agentless}) based on twelve advanced LLMs (i.e., Claude-3.7-Sonnet, Claude-3.5-Sonnet, DeepSeek-R1, DeepSeek-V3, Doubao-1.5-pro, Doubao-1.5-thinking, GPT-4o, OpenAI-o1, OpenAI-o3-mini-high, Llama-4-Maverick, Qwen2.5-72B-Instruct, Gemini-2.5-Pro). 
It is worth noting that these agent frameworks were originally designed for Python projects, as used in SWE-bench~\cite{jimenez2024swebench}. 
To support the multilingual setting of Multi-SWE-bench, \citet{zan2025multiswebench} adapted these frameworks (e.g., \path{Agentless}\footnote{\url{https://github.com/OpenAutoCoder/Agentless}} $\rightarrow$ \path{MagentLess}\footnote{\url{https://github.com/multi-swe-bench/MagentLess}}).
As these adaptations do not alter the core design principles of the original agent frameworks, we refer to them using their original names throughout this paper.

\subsection{Data Filtering}
% https://capitalizemytitle.com/
\subsubsection{Issue-Based Data Filtering}
As this study investigates tangled refactorings in software issue resolution, it is necessary to ensure that the analyzed issues are not primarily concerned with refactoring itself, but instead involve tasks such as bug fixing or feature request.
To this end, we manually examined the titles and descriptions of the 128 software issues in Multi-SWE-bench to determine whether they explicitly requested refactoring operations and to validate the issue labels provided in the dataset (i.e., \textit{bug fix}, \textit{new feature}, and \textit{feature optimization}).
Our examination confirmed that the issues are consistent with these labels and are not primarily intended for refactoring.
Consequently, any refactoring operations identified in patches generated by coding agents are likely unintentional refactorings that constitute tangled commits, where refactoring modifications are combined with functional changes (e.g., bug fixes or feature updates) within the same commit.

\subsubsection{Patch-Based Data Filtering}
To detect refactoring operations from both human-written and agent-generated patches (4,736 in total), we first apply each patch to the corresponding repository at its base commit.
To remain consistent with the evaluation framework of Multi-SWE-bench,\footnote{\url{https://github.com/multi-swe-bench/multi-swe-bench}} we adopt the same two-stage patch application procedure.
Specifically, we first attempt to apply each patch using ``\texttt{git apply --whitespace=nowarn}''. 
If this step fails (e.g., due to minor whitespace inconsistencies or mismatched context lines), we fall back to the Unix command (i.e., ``\texttt{patch --batch --fuzz=5 -p1}''), which allows a fuzz factor of up to five lines of contextual deviation. 
This two-stage process mirrors the procedure used in the Multi-SWE-bench evaluation tool, ensuring consistency with the benchmark's validation pipeline.

Among the 4,608 agent-generated patches, 3,693 (80.1\%) were successfully produced and applied valid patches, 441 (9.6\%) resulted in empty patches, and 474 (10.3\%) failed to apply. 
The empty patches correspond to cases in which the coding agent did not generate valid patch content for the given software issue. 
The application failures occur when the diff hunks cannot be matched to the target source files, even with the relaxed fuzz factor. 
Common causes include references to non-existent file paths, incorrect line-number offsets exceeding the fuzz tolerance, or malformed diff syntax.
In the remainder of this study, we focus on the 128 human-written patches and the 3,693 successfully applied agent-generated patches.

\subsection{Tangled Refactoring Detection} 
To automatically detect and classify tangled refactorings within patches, we employ the widely-used refactoring detection tool RefactoringMiner~\cite{Alikhanifard:TOSEM:2024:RefactoringMiner3.0}, which has been extensively adopted in empirical software engineering research for mining refactoring histories in Java repositories~\cite{refactoringMiner1, refactoringMiner2,buggysideofrefactoring}. 
We select RefactoringMiner due to its higher coverage and accuracy compared with alternative tools such as RefDiff~\cite{refdiff}. 
Specifically, RefactoringMiner leverages an AST-based statement matching algorithm that enables the detection of more than 100 distinct refactoring types, ranging from high-level structural changes (e.g., \textit{Move Method} and \textit{Move Class}) to fine-grained operations (e.g., \textit{Extract Method} and \textit{Rename Variable}). 
In contrast, RefDiff supports only 11 refactoring types with approximately 96\% precision and 80\% recall.

Specifically, we apply RefactoringMiner (version 3.0) to both the 3,693 agent-generated patches and the 128 human-written golden patches to detect tangled refactoring operations.
During the analysis, we identified two abnormal patches generated by the coding agent OpenHands based on Gemini 2.5 Pro. 
These patches performed project-wide version rollbacks in the \textit{jackson-databind} repository, downgrading the project from 2.17.0-SNAPSHOT to 2.15.3 and 2.16.0-SNAPSHOT to 2.15.2, respectively. 
As a result, they modified 476 and 111 files, including large portions of the test suite. 
These rollbacks were incorrectly identified as 1,122 refactoring operations (e.g., 653 removals of JUnit 5 @Test annotations caused by the framework downgrade). 
To avoid skewing the analysis, we further excluded these two abnormal patches, leaving 3,691 agent-generated patches for the subsequent analysis.
In total, we detected 2,429 tangled refactoring instances from the 3,691 agent-generated patches, covering 73 unique refactoring types. 
At the patch level, 791 out of 3,691 patches (21.43\%) contain at least one tangled refactoring. 
For the human-written patches, we identified 224 tangled refactoring instances spanning 46 distinct refactoring types, with 47 out of 128 patches (36.72\%) containing at least one tangled refactoring.

\begin{table}[t]
\centering
\tabcolsep=3mm
\caption{Tangled refactoring presence: agent-generated vs. human-written patches.}
\label{tab:refactoring_quadrants}
\begin{tabular}{lccccc}
    \toprule
    \textbf{Metric} & \textbf{\makecell{Both\\ Refactors}} & \textbf{\makecell{Agent\\ Refactors}} & \textbf{\makecell{Golden\\ Refactors}} & \textbf{\makecell{Neither\\ Refactors}} & \textbf{Total}  \\
    \midrule
    \textbf{Count (Ratio)} & 398 (10.78\%) & 393 (10.65\%) & 942 (25.52\%) & 1,958 (53.05\%) & 3,691 \\
    \bottomrule
    \multicolumn{6}{l}{\footnotesize * Both Refactors: both the agent-generated patch and the golden patch contain at least one tangled refactoring;} \\
    \multicolumn{6}{l}{\footnotesize * Agent Refactors: only the agent-generated patch contains tangled refactoring;} \\ 
    \multicolumn{6}{l}{\footnotesize * Golden Refactors: only the golden patch contains tangled refactoring;} \\
    \multicolumn{6}{l}{\footnotesize * Neither Refactors: neither patch contains refactoring.}
\end{tabular}
\end{table}

Table~\ref{tab:refactoring_quadrants} presents the distribution of tangled refactorings in agent-generated patches and their corresponding human-written golden patches, organized into four categories. 
Specifically, the categories include: 
(1) \textit{Both Refactors}, where both the agent-generated patch and the golden patch contain at least one tangled refactoring; 
(2) \textit{Agent Refactors}, where only the agent-generated patch contains tangled refactoring; 
(3) \textit{Golden Refactors}, where only the golden patch contains tangled refactoring; 
and (4) \textit{Neither Refactors}, where neither patch contains refactoring.
Based on the analysis of 3,691 agent-generated patches, we observe that 1,733 patches (46.95\%) contain at least one tangled refactoring, indicating that such refactorings are prevalent in agent-based issue resolution and highlighting the importance of studying this phenomenon. 
Among these cases, 398 patches (10.78\%) fall into the \textit{Both Refactors} category, where both coding agents and human developers perform refactoring while resolving the issue, suggesting that certain maintenance tasks inherently require structural improvements. 
In 393 patches (10.65\%), categorized as \textit{Agent Refactors}, the coding agent performs refactoring while the human-written golden patch does not, indicating that coding agents may reorganize surrounding code even when a minimal fix would suffice. 
Conversely, 942 patches (25.52\%), categorized as \textit{Golden Refactors}, involve refactoring in the human-written golden patch but not in the agent-generated patch, suggesting that human developers are more inclined to improve code structure during maintenance. 
Finally, 1,958 patches (53.05\%) fall into the \textit{Neither Refactors} category, where neither agents nor human developers perform refactoring.

\section{Empirical Results}
% In this section, We provide the approach and results for each of our research questions. 

\begin{table*}[t]
\centering
\footnotesize
\caption{Differences in tangled refactorings between coding agents and human developers in terms of Tangled Refactoring Ratio, Density, and Types, respectively. Note that \textbf{bold} and \underline{underlined} values indicate best-performing and second-best results, respectively.}
\label{tab:combined_refactoring_stats}
\tabcolsep=1.9mm
\normalsize
\renewcommand{\arraystretch}{0.8}
\begin{threeparttable}
% \resizebox{\textwidth}{!}{
    \begin{tabular}{l ccc} 
    \toprule
    \multirow{1}{*}{\textbf{Category}}
    & \textbf{\makecell{Tangled \\ Refactoring Ratio\tnote{1}}}
    & \textbf{\makecell{Tangled \\ Refactoring Density\tnote{2}}}
    & \textbf{\makecell{Tangled \\ Refactoring Types\tnote{3}}} \\
    \midrule
    \multicolumn{4}{l}{\cellcolor{gray!30}{\textbf{Aggregated Results Grouped by Agent Frameworks}}} \\ 
    \midrule
    \textbf{SWE-agent}    & \textbf{25.85} ($\frac{371}{1,435}$) & \textbf{1.19} ($\frac{1,703}{1,435}$) & \textbf{73}  \\
    \textbf{Agentless}   & \underline{22.04} ($\frac{266}{1,207}$) & \underline{0.38} ($\frac{454}{1,207}$) & 35  \\
    \textbf{OpenHands}    & 14.68 ($\frac{154}{1,049}$) & 0.26 ($\frac{272}{1,049}$) & \underline{36}  \\
    \midrule
    \multicolumn{4}{l}{\cellcolor{gray!30}{\textbf{Aggregated Results Grouped by Base LLMs}}} \\ 
    \midrule
    \textbf{Claude-3.5-Sonnet}  & \underline{28.48} ($\frac{88}{309}$) & 0.74 ($\frac{230}{309}$) & \textbf{50} \\
    \textbf{Claude-3.7-Sonnet}       & 24.91 ($\frac{73}{293}$) & 0.57 ($\frac{166}{293}$) & 26 \\
    \textbf{DeepSeek-R1}             & 19.33 ($\frac{58}{300}$) & 0.61 ($\frac{183}{300}$) & 34 \\
    \textbf{DeepSeek-V3}             & 20.20 ($\frac{62}{307}$) & 0.43 ($\frac{131}{307}$) & 26 \\
    \textbf{Doubao-1.5-pro}          & 11.46 ($\frac{29}{253}$) & 0.17 ($\frac{42}{253}$)  & 12 \\
    \textbf{Doubao-1.5-thinking}     & 24.92 ($\frac{82}{329}$) & \textbf{1.12} ($\frac{370}{329}$) & \underline{49} \\
    \textbf{GPT-4o-1120}             & 20.82 ($\frac{66}{317}$) & \underline{0.98} ($\frac{310}{317}$) & 34 \\
    \textbf{OpenAI-o1}               & 17.83 ($\frac{56}{314}$) & 0.57 ($\frac{178}{314}$) & 31 \\
    \textbf{OpenAI-o3-mini-high}     & 21.51 ($\frac{57}{265}$) & 0.45 ($\frac{119}{265}$) & 27 \\
    \textbf{Llama-4-Maverick}        & 20.88 ($\frac{76}{364}$) & 0.57 ($\frac{207}{364}$) & 39 \\
    \textbf{Qwen2.5-72B-Instruct}    & 15.20 ($\frac{45}{296}$) & 0.77 ($\frac{227}{296}$) & 34 \\
    \textbf{Gemini-2.5-Pro}          & \textbf{28.78} ($\frac{99}{344}$) & 0.77 ($\frac{266}{344}$) & 37 \\
    \midrule
    \multicolumn{4}{l}{\cellcolor{gray!30}{\textbf{Overall Results of Coding Agents and Human Developers}}} \\ 
    \midrule
    \textbf{Coding Agents}   & \underline{21.43} ($\frac{791}{3,691}$) & \underline{0.66} ($\frac{2,429}{3,691}$) & \textbf{73}  \\
    \textbf{Human}   & \textbf{36.72} ($\frac{47}{128}$) & \textbf{1.75} ($\frac{224}{128}$) & \underline{46}  \\
    \bottomrule
    \end{tabular}

    \begin{tablenotes}[flushleft]
        \footnotesize
        \setlength{\itemsep}{0pt}
        \item[1] \textbf{Tangled Refactoring Ratio (\%)} = (\# Patches with at least one tangled refactoring) / (\# Patches);
        \item[2] \textbf{Tangled Refactoring Density} = (\# Tangled Refactorings) / (\# Patches);
        \item[3] \textbf{Tangled Refactoring Types} = (\# Types of Tangled Refactorings).
    \end{tablenotes}
\end{threeparttable}
\end{table*}

\subsection{RQ1: Differences in Tangled Refactorings between Coding Agents and Human Developers}

% coding agents differ from human developers in their use of refactoring for issue resolution?}

\subsubsection{Approach} 
In this research question, we conduct a systematic comparison of the tangled refactorings employed by coding agents and human developers in patches created for issue resolution. 
Specifically, we examine the differences between these two approaches by analyzing three metrics: 
(1) \textbf{Tangled Refactoring Ratio}, defined as the proportion of patches that contain at least one tangled refactoring;
(2) \textbf{Tangled Refactoring Density}, defined as the average number of tangled refactorings applied per patch; 
and (3) \textbf{Tangled Refactoring Types}, referring to the total number of refactoring types identified in the patches.

\subsubsection{Results} 
\begin{figure}[t]
    \centering
    \includegraphics[width=.9\linewidth]{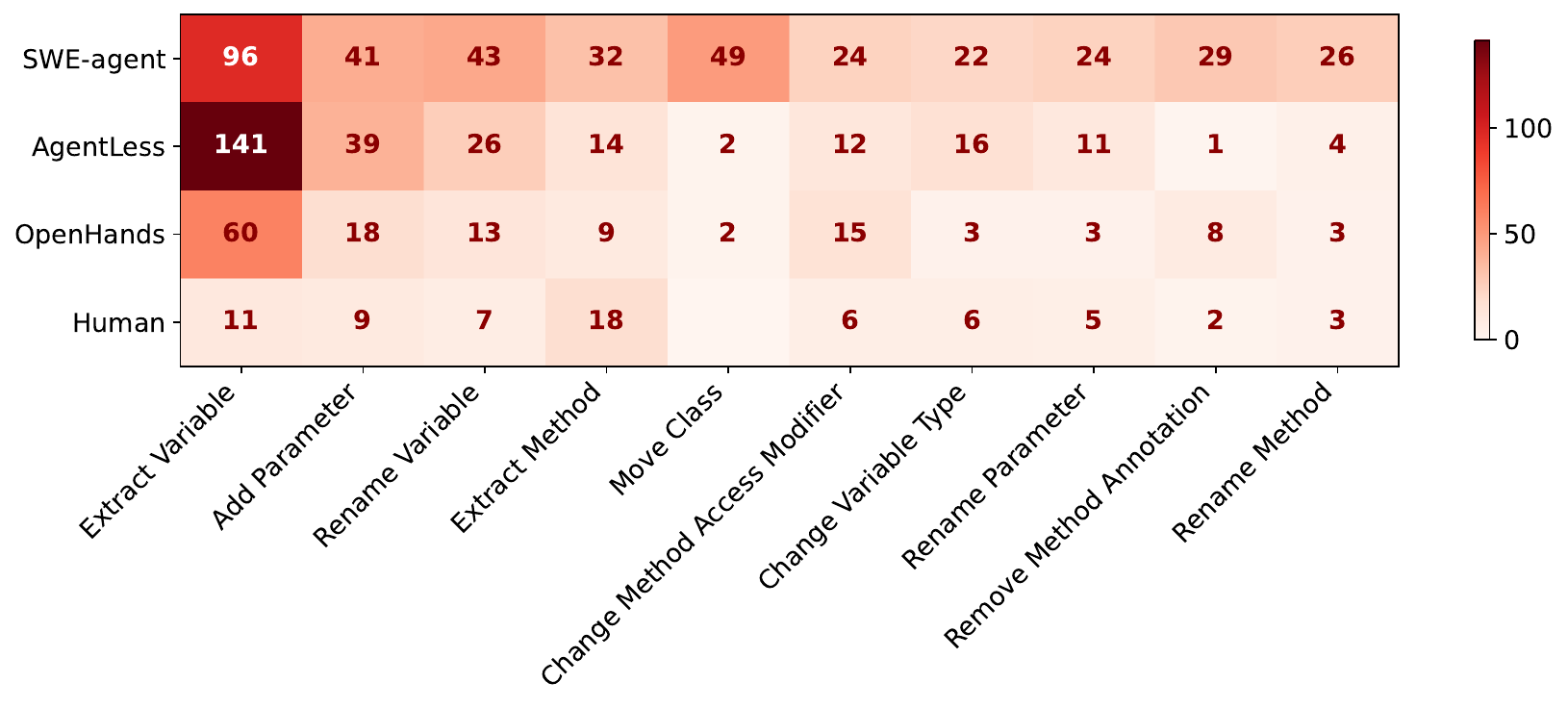} 
    \caption{Distribution of the top-10 tangled refactoring types by agent framework (aggregated across 3 agent frameworks). Each cell denotes the number of patches with at least one tangled refactoring.}
    \label{fig:ref_agent_heatmap}
\end{figure}

\begin{figure}[t]
    \centering
    \includegraphics[width=\linewidth]{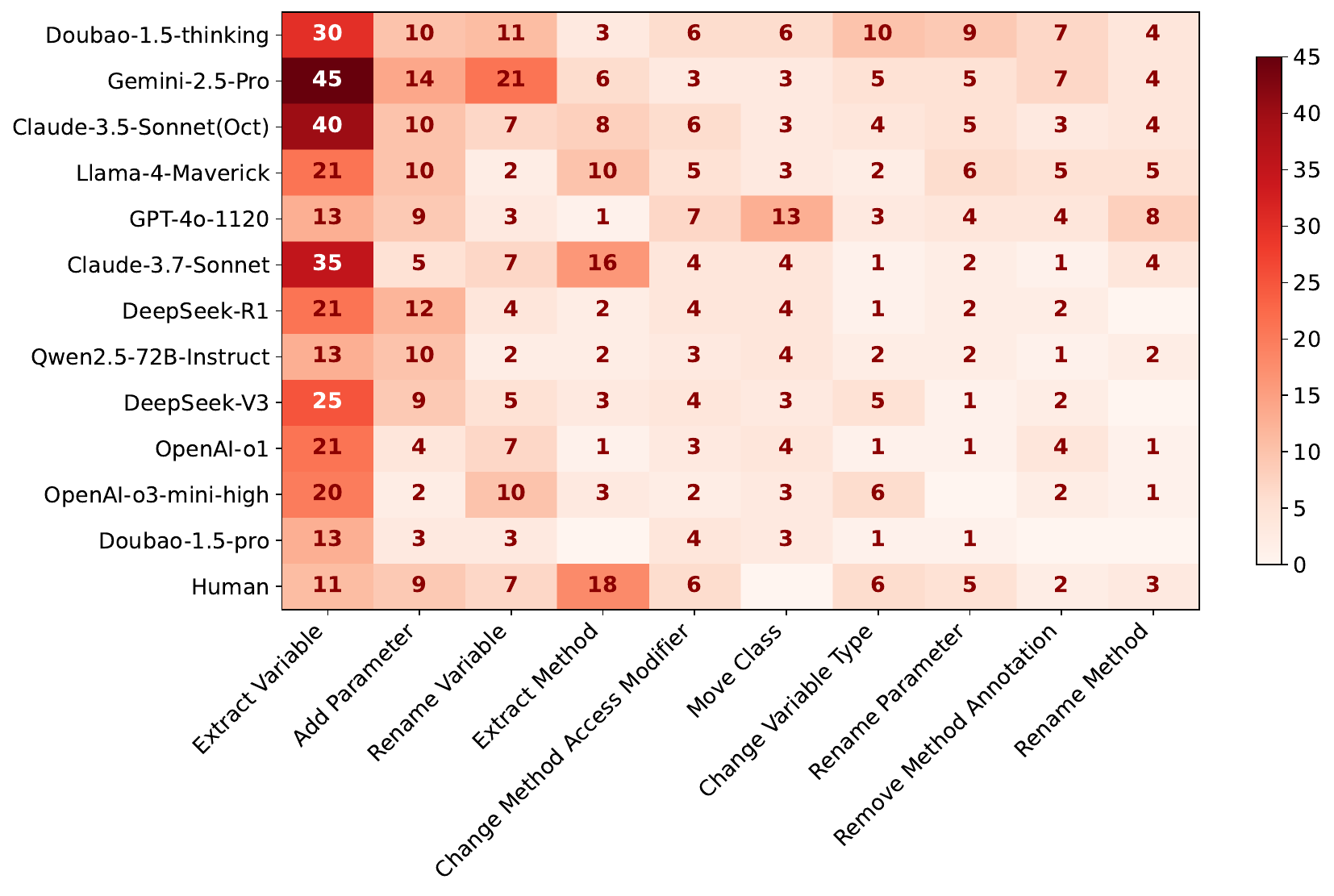} 
    \caption{Distribution of the top-10 tangled refactoring types by base LLM (aggregated across 12 LLMs). Each cell denotes the number of patches with at least one tangled refactoring.}
    % \wang{x, y-axis hard to read}
    \label{fig:ref_llm_heatmap}
\end{figure}

Table~\ref{tab:combined_refactoring_stats} presents the comparative results between coding agents and human developers, along with aggregated refactoring statistics grouped by the three agent frameworks and the twelve base LLMs, respectively. 
In the table, bold values indicate the best-performing result, while underlined values denote the second-best result. 
In addition, Figures~\ref{fig:ref_agent_heatmap} and ~\ref{fig:ref_llm_heatmap} further illustrate the distribution of the top-10 tangled refactoring types across the three agent frameworks and the twelve base LLMs, respectively.

\textbf{Metric I. Tangled Refactoring Ratio.} 
The results indicate that human developers perform refactoring more frequently than coding agents during issue resolution.
As shown in Table~\ref{tab:combined_refactoring_stats}, 36.72\% of human-written patches contain at least one tangled refactoring, compared to only 21.43\% of agent-generated patches.
This difference suggests that human developers are more inclined to reorganize or improve code structure even while addressing concrete issues.

Across different agent frameworks, OpenHands exhibits the lowest proportion of tangled refactorings, with only 14.68\% of its patches involving refactoring. 
In contrast, SWE-agent and Agentless perform refactoring in 25.85\% and 22.04\% of patches, respectively. 
These discrepancies may be attributed to differences in agent design.
Specifically, Agentless follows a workflow-based paradigm~\cite{ceka2025understandingsoftwareengineeringagents}, where its framework executes a pre-defined pipeline of subtasks (e.g., localization, patch generation, and patch validation). 
In contrast, both SWE-agent and OpenHands adopt a more open-ended process~\cite{ceka2025understandingsoftwareengineeringagents}, granting greater autonomy through flexible tool interactions. 
Such differences in interaction design may influence the extent to which agents introduce refactoring during issue resolution.

Across different base LLMs, Gemini-2.5-Pro and Claude-3.5-Sonnet exhibit the highest proportion of tangled refactorings, with 28.78\% and 28.48\% of their patches involving refactoring, respectively. 
These models are representative of reasoning models (which employ the chain-of-thought before answering) and general-purpose models (referred to as ``general models'' in the rest of the paper).
Nonetheless, even these advanced LLMs still perform fewer tangled refactorings than human developers (36.72\%). 
Furthermore, models from the same provider may exhibit substantially different behaviors, despite the speculation that they may share a significant portion of their training corpora. 
For instance, ByteDance's general model Doubao-1.5-pro achieves the lowest refactoring ratio (11.46\%), whereas its reasoning counterpart Doubao-1.5-thinking reaches 24.92\%. 
In contrast, the refactoring ratio of DeepSeek-R1 (19.33\%) is comparable to that of its general model counterpart DeepSeek-V3 (20.20\%).

\textbf{Metric II. Tangled Refactoring Density.} 
The results further show that human developers perform substantially more refactoring than coding agents.
As reported in Table~\ref{tab:combined_refactoring_stats}, human-written patches contain, on average, 1.75 tangled refactoring operations per patch, whereas agent-generated patches contain only 0.66 on average. 
This observation is consistent with the findings for the tangled refactoring ratio metric, reinforcing that human developers are generally more inclined to apply refactoring during issue resolution.

Across different agent frameworks, SWE-agent exhibits the highest tangled refactoring density, with an average of 1.19 refactoring operations per patch. 
In contrast, Agentless, despite having 22.04\% of its patches include tangled refactorings, achieves a relatively low density of 0.38 refactoring operations per patch, indicating that its refactoring actions are less intensive when they occur.
Besides, OpenHands shows the lowest tangled refactoring density, with an average of 0.26 refactoring operations per patch.

Across different base LLMs, Doubao-1.5-thinking exhibits the highest tangled refactoring density, with 1.12 refactoring operations per patch, while all other models perform fewer than one refactoring operation per patch on average.
Notably, not all agent-generated patches successfully resolve the target issues.
This suggests that coding agents may underutilize refactoring compared to human developers, potentially limiting their effectiveness. 
We further investigate the impact of tangled refactorings on issue resolution performance in RQ2 (Section~\ref{sec: 2}).

\textbf{Metric III. Tangled Refactoring Types.} 
The results indicate that coding agents exhibit a broader diversity of tangled refactoring types than human developers. 
As shown in Table~\ref{tab:combined_refactoring_stats}, agent-generated patches cover 73 distinct refactoring types, whereas human-written patches involve only 46 types. 
This suggests that coding agents explore a wider range of refactoring operations than human developers, although not all of them are necessarily required for effective issue resolution.

Across different agent frameworks, SWE-agent covers all 73 distinct tangled refactoring types observed in the dataset, representing the full spectrum of refactoring behaviors exhibited by coding agents. 
In contrast, Agentless (35 types) and OpenHands (36 types) exhibit considerably fewer unique tangled refactoring types than human developers (46 types). 
This disparity suggests that refactoring diversity is strongly influenced by the agent framework design: frameworks with greater autonomy and fewer constraints (e.g., SWE-agent) tend to explore a broader set of refactoring types, whereas more structured, workflow-based approaches (e.g., Agentless) \cite{ceka2025understandingsoftwareengineeringagents} result in a more limited set of transformations.

Across different base LLMs, only Claude-3.5-Sonnet (50 types) and Doubao-1.5-thinking (49 types) exceed the diversity observed in human-written patches (46 types), while all other LLMs fall below this level. 
This observation suggests that the diversity of tangled refactoring types is more strongly shaped by the agent framework than by the underlying LLM. In contrast, ByteDance's general model Doubao-1.5-pro carried out only 12 distinct tangled refactoring types. This is consistent with the findings for the tangled refactoring ratio metric, where it achieves the lowest refactoring ratio.

% \tao{how did we switch from base LLMs to coding agent}In particular, SWE-agent alone covers all 73 observed refactoring types, likely due to its higher degree of autonomy, which encourages exploration of diverse code transformations. 
% In contrast, Agentless, which follows a fixed three-phase pipeline (localization, patch generation, and patch validation), produces a more constrained set of refactoring types.\tao{this is also repetitive with the end of the last paragraph}

Regarding the specific types of tangled refactorings, all agent frameworks exhibit broadly similar patterns, as illustrated in Figure~\ref{fig:ref_agent_heatmap}. 
In particular, \textit{Extract Variable} is consistently and extensively applied to simplify complex expressions and improve code readability.
Another frequently observed refactoring is \textit{Add Parameter}, which introduces additional inputs to existing methods to support functional changes.
Beyond these shared patterns, SWE-agent more frequently applies \textit{Move Class}, relocating classes to more appropriate packages, typically to improve modularity. 
In contrast, human developers are more inclined to apply \textit{Extract Method} during issue resolution, decomposing code fragments into separate, well-named methods to enhance readability, reduce duplication, and improve maintainability.

As shown in Figure~\ref{fig:ref_llm_heatmap}, Gemini-2.5-Pro and Claude-3.5-Sonnet are the two base LLMs that most frequently apply \textit{Extract Variable} refactoring (45 and 40 patches, respectively).
Other LLMs exhibit similar patterns, with \textit{Extract Variable} consistently being the most dominant refactoring type. 
This may suggest that LLMs have effectively learned this refactoring pattern from their training data. 
Prior empirical studies~\cite{Tsantalis:TSE:2020:RefactoringMiner2.0} further support this observation, reporting \textit{Extract Variable} as one of the most frequent refactoring types (e.g., the second most common in a benchmark of 536 commits from 185 open-source projects), while \textit{Move Class} ranks as the most frequent.
Notably, human developers exhibit a distinct pattern: \textit{Extract Method} (18 patches) is their most frequently applied tangled refactoring type, surpassing \textit{Extract Variable} (11 patches).
This difference indicates that human developers place greater emphasis on method-level decomposition during issue resolution, whereas coding agents tend to favor more localized, expression-level transformations.

\begin{tcolorbox}[colframe=black,colback=gray!20]
\textbf{RQ1 Summary.}
% Among agent frameworks, SWE-agent leads in refactoring ratio density, and type diversity; among base LLMs, Gemini-2.5-Pro and Claude-3.5-Sonnet(Oct) achieve the highest refactoring ratios, while Doubao-1.5-thinking exhibits the highest density.
Human developers apply tangled refactoring both more frequently and more intensively than coding agents when resolving real-world issues. 
Although coding agents collectively exhibit a broader range of refactoring types than humans, this diversity is primarily driven by agent framework design (notably SWE-agent) rather than the choice of base LLM. 
At the operation level, agent refactoring is dominated by lightweight transformations such as \textit{Extract Variable}, whereas human developers more frequently employ method-level refactorings (particularly \textit{Extract Method}). 
This contrast suggests that human developers place greater emphasis on improving code modularity and readability during issue resolution.
\end{tcolorbox}

\subsection{RQ2: Impact of Tangled Refactorings on Coding Agents' Issue Resolution}
\label{sec: 2}

\subsubsection{Approach}
To investigate the impact of tangled refactorings on coding agents' effectiveness in issue resolution, we fit logistic regression models~\cite{hosmer2013applied} with two binary outcome variables (\textbf{Compilability} and \textbf{Correctness}). 
These logistic regression models estimate the marginal effects of refactoring-related factors while controlling for task-, agent-, and patch-level characteristics. 
Table~\ref{tab:rq2_variables} summarizes all studied variables, including independent, control, and dependent variables, along with their descriptions and data types.

\begin{table*}[t]
  \centering
  \caption{The studied variables of logistic regression models in RQ2.}
  \label{tab:rq2_variables}
  \footnotesize
  \renewcommand{\arraystretch}{1.2}
  \definecolor{LightGray}{gray}{0.85}

  \begin{tabularx}{\textwidth}{l X l}
    \toprule
    \textbf{Variable} & \textbf{Description} & \textbf{Type} \\
    \midrule

    \rowcolor{LightGray} \multicolumn{3}{l}{\textbf{Independent Variables}} \\
    \midrule
    
    Refactoring Presence & Whether a patch contains at least one refactoring operation, i.e., \path{has_refactoring} & Binary \\
    % Refactoring Action~\cite{refactoring_impact_on_regression} & Set of binary indicators for refactoring actions, i.e., \path{has_add}, \path{has_remove}, and \path{has_adjust}. & Binary (Set) \\
    % Refactoring Scope~\cite{doesrefactoringbreaktests} & Set of binary indicators for refactoring scope, i.e., \path{has_method}, \path{has_class}, and \path{has_local_variable}. & Binary (Set) \\

    \makecell[l]{Refactoring Action~\cite{refactoring_impact_on_regression} \\ \& Scope~\cite{doesrefactoringbreaktests}} & \makecell[l]{Refactoring action comprises binary indicators for adding, removing, or \\ adjusting code (i.e., \path{has_add}, \path{has_remove}, and \path{has_adjust}), while \\ refactoring scope indicates the refactoring level (i.e., \path{has_method}, \\ \path{has_class}, and \path{has_local_variable}).} & Binary (Set) \\
    
    \midrule
    \rowcolor{LightGray} \multicolumn{3}{l}{\textbf{Control Variables: Task Characteristics~\cite{zan2025multiswebench}}} \\
    \midrule
    
    Task Difficulty & The difficulty level of the task, i.e., \path{Easy}, \path{Medium}, or \path{Hard}. & Categorical \\
    Issue Type & The category of the issue, i.e., \path{bug_fix}, \path{new_feature} or  \path{feature_optimization}. & Categorical \\
    Issue Length & The token length of the issue description. & Numeric \\
    
    \midrule
    \rowcolor{LightGray} \multicolumn{3}{l}{\textbf{Control Variables: Agent Characteristics~\cite{zan2025multiswebench}}} \\
    \midrule
    
    Base Model & The base LLM used by the coding agent, e.g., Claude-3.7-Sonnet, OpenAI-o1, or DeepSeek-V3. & Categorical \\
    Agent Framework & The employed agent framework, i.e., \path{SWE-agent}, \path{OpenHands}, or \path{Agentless}. & Categorical \\

    \midrule
    \rowcolor{LightGray} \multicolumn{3}{l}{\textbf{Control Variables: Patch Characteristics~\cite{MAGIS}}} \\
    \midrule
    
    Modified Lines & Total number of lines changed in the generated patch. & Numeric \\
    Modified Files & Number of files modified in the generated patch. & Numeric \\
    File Coverage & File-level overlap ratio between generated and reference patches. & Numeric [0--1] \\
    Line Coverage & Line-level overlap ratio between generated and reference patches. & Numeric [0--1] \\

    \midrule
    \rowcolor{LightGray} \multicolumn{3}{l}{\textbf{Dependent Variables}} \\
    \midrule
    
    Compatibility & Whether the generated patch compiles successfully. & Binary \\
    Correctness & Whether the generated patch passes all gold tests and thus resolves the issue. & Binary \\

    \bottomrule
  \end{tabularx}
\end{table*}

\textbf{Independent Variables.}
Our primary independent variable is \path{has_refactoring}, a binary indicator capturing whether a generated patch contains any tangled refactoring.
To further examine the effects of refactoring characteristics, we decompose tangled refactorings along two orthogonal dimensions commonly adopted in prior refactoring studies:
(i) \textit{Refactoring Action}~\cite{refactoring_impact_on_regression}, which characterizes whether code elements are added, removed, or adjusted in refactoring practice; and
(ii) \textit{Refactoring Scope}~\cite{doesrefactoringbreaktests}, which captures the structural level at which the refactoring operates (i.e., class, method, or local variable) in refactoring practice.
These characteristics are encoded as sets of binary dummy variables (e.g., \path{has_add}). 
This decomposition enables us to disentangle the effects of refactoring actions from those of refactoring scopes, which may differ in their influence on coding agents' effectiveness in issue resolution.

\textbf{Control Variables.}
To isolate the effect of refactoring, we control for a range of potential confounding factors identified in prior work~\cite{MAGIS, zan2025multiswebench}. 
Specifically, task difficulty and issue type capture variation in problem complexity, while issue length reflects the amount of information provided in the issue description.
We further include the base LLM and agent framework to account for differences in coding agent capabilities.
At the patch level, we control for modified lines (i.e., the total number of added and deleted lines) and modified files (i.e., the number of affected files), as larger and more dispersed patches are generally more error-prone~\cite{nagappan2005use,hassan2009predicting}.
In addition, we introduce two coverage-based metrics to quantify the structural alignment between agent-generated patches and human-written (golden) patches:
% \tao{use the same variable name as original paper for formula, and using original formular; this formula is not the same as: MAGIS: LLM-based multi-agent framework for GitHub issue ReSolution } 
\begin{enumerate}
    \item \textit{File Coverage.} Let $F_{\text{human}}$ denote the set of files modified in the golden patch and $F_{\text{agent}}$ denote the set of files modified in the agent-generated patch.
    File coverage is calculated as,
    \[
    \text{File Coverage} =
    \frac{|F_{\text{agent}} \cap F_{\text{human}}|}{|F_{\text{human}}|}
    \]

    \item \textit{Line Coverage.} 
    For each file $f \in F_{\text{human}}$, let $L_{\text{human}}^{f}$ and $L_{\text{agent}}^{f}$ denote the sets of modified line numbers in the golden and generated patches, respectively.
    Line coverage is computed as,
    \[
    \text{Line Coverage} =
    \frac{\sum_{f \in F_{\text{human}}} |L_{\text{agent}}^{f} \cap L_{\text{human}}^{f}|}
    {\sum_{f \in F_{\text{human}}} |L_{\text{human}}^{f}|}
    \]
    When multiple hunks exist, the line sets are defined as the union of all modified line ranges within each file.
\end{enumerate}

\textbf{Data Preprocessing and Multicollinearity Diagnosis.}
Prior to fitting the logistic regression model, we perform a series of diagnostic checks to ensure statistical validity, following established practices in empirical software engineering.
Our analysis is based on 3,691 valid instances, corresponding to patches that can be successfully applied to their associated codebases as explained in Section~\ref{sec:dp}.
To address skewness in numeric variables, we apply the logarithmic transformation (log1p) to right-skewed features, including \path{modified_lines} (skewness $\approx$ 8.24), \path{modified_files} (skewness $\approx$ 8.41), and \path{issue_length} (skewness $\approx$ 6.5).
Categorical variables (i.e., task difficulty, issue type, base model, and agent framework) are one-hot encoded with a drop-first reference level to avoid the dummy variable trap. 
To assess potential multicollinearity, we compute the variance inflation factors (VIF) for all explanatory variables~\cite{kim2019multicollinearity}.
The maximum VIF observed is below 3 (e.g., \path{task_difficulty_Medium} $\approx$ 2.96 and \path{modified_lines_log_scaled} $\approx$ 2.23), which is well below the commonly accepted threshold of 5, indicating no severe multicollinearity~\cite{VIF}. 
Accordingly, all control variables are retained in the final models.

\textbf{Model Fit.}
To analyze the effects of refactoring on the two dependent variables (\textit{Compilability} and \textit{Correctness}), we construct four distinct logistic regression models, corresponding to two sets of independent variables (i.e., \textit{Refactoring Presence} and \textit{Refactoring Action \& Scope}) and two dependent variables.
We report \textit{Odds Ratios} (\textit{OR}) to quantify the direction and magnitude of the associations, where $OR > 1$ indicates a positive association and $OR < 1$ indicates a negative association. 
Statistical significance is determined using \path{p-values}. 
To assess the goodness-of-fit of the model, we report the \textit{Adjusted McFadden's pseudo-$R^2$}, which provides a conservative measure by accounting for the number of independent variables.

\subsubsection{Results}
Table~\ref{tab:rq2_effects} reports the estimated \textit{Odds Ratios} (\textit{OR}), $p$-values, and variable-level instance number ($N$) for all independent variables.
Across the four logistic regression models, the adjusted McFadden's pseudo-$R^2$ values are 0.310 (\textit{Refactoring Presence} $\rightarrow$ \textit{Correctness}), 0.181 (\textit{Refactoring Presence} $\rightarrow$ \textit{Compilability}), 0.323 (\textit{Refactoring Action \& Scope} $\rightarrow$ \textit{Correctness}), and 0.193 (\textit{Refactoring Action \& Scope} $\rightarrow$ \textit{Compilability}). 
The relatively lower model fit for \textit{Compilability} (0.181 and 0.193), compared to \textit{Correctness} (0.310 and 0.323), is expected, as \textit{Compilability} is more significantly influenced by additional factors (e.g., build systems, dependencies, and environment configurations) that are not fully captured by our predictors. 
Nevertheless, adjusted pseudo-$R^2$ values around 0.18$\sim$0.19 remain acceptable and informative for logistic regression models in empirical software engineering studies~\cite{hosmer2013applied}.

\textbf{Impact of Refactoring Presence on Coding Agents' Effectiveness.} 
For \textit{Compilability}, the presence of tangled refactorings in agent-generated patches is significantly negatively associated with successful compilation ($OR = 0.42$, $p = 1.76e\text{-}18$), indicating a reduction of approximately 58\% in the odds of successful compilation after controlling for task-, agent-, and patch-level factors. 
In contrast, for \textit{Correctness}, the refactoring presence does not exhibit a statistically significant effect ($OR = 1.09$, $p = 0.543$), suggesting that the inclusion of tangled refactorings shows no statistically significant association with issue resolution success.

\textbf{Impact of Refactoring Action and Scope on Coding Agents' Effectiveness.}
The results further reveal heterogeneous effects across different refactoring characteristics.
Removal-related (\path{has_remove}) refactoring operations (e.g., \emph{Remove Parameter} and \emph{Remove Method Annotation}) are positively associated with both \textit{Correctness} ($OR = 4.56$, $p = 0.002$) and \textit{Compilability} ($OR = 2.50$, $p = 0.006$), suggesting that eliminating unnecessary or problematic code can improve agent effectiveness.
In contrast, adjust-related (\path{has_adjust}) refactoring operations (e.g., \emph{Rename Variable} and \emph{Extract Method}) show a negative association with \textit{Compilability} ($OR = 0.55$, $p = 0.069$), although this effect does not reach statistical significance at the 0.05 level.
Regarding refactoring scope, method-level refactorings (\path{has_method}) are significantly negatively associated with both \textit{Correctness} ($OR = 0.22$, $p = 0.003$) and \textit{Compilability} ($OR = 0.23$, $p = 4.59e\text{-}06$), indicating that structural modifications at the method level are correlated with lower effectiveness of coding agents in issue resolution.
In contrast, class-level and local-variable-level refactorings do not exhibit statistically significant effects ($p > 0.05$).
Overall, these results suggest that method-level refactorings are consistently associated with reduced effectiveness, whereas removal-related refactorings tend to improve coding agents' performance in issue resolution.

\begin{table*}[t]
  \centering
  \caption{Effects of refactoring-related variables on task outcomes (logistic regression). \textit{OR} denotes \textit{Odds Ratio}. Significance: $^{*}$ $p < 0.05$, $^{**}$ $p < 0.01$, $^{***}$ $p < 0.001$.}
  \label{tab:rq2_effects}
  \tabcolsep=3.6mm
  \normalsize
  % \footnotesize
  \renewcommand{\arraystretch}{1.2}
  \definecolor{LightGray}{gray}{0.85}

  \begin{tabularx}{\textwidth}{l c X X}
    \toprule
    \textbf{Variable} & \textbf{$N$} & \textbf{Correctness (OR; $p$)} & \textbf{Compilability (OR; $p$)} \\
    \midrule

    \rowcolor{LightGray} \multicolumn{4}{l}{\textbf{Refactoring Presence}} \\
    \midrule

    \path{has_refactoring} & 791 & 1.09; $p = 0.543$ & 0.42; $p = 1.76\mathrm{e}{-18}^{***}$ \\

    \midrule
    \rowcolor{LightGray} \multicolumn{4}{l}{\textbf{Refactoring Dimension: Action}} \\
    \midrule

    \path{has_add} & 173 & 0.43; $p = 0.111$ & 0.88; $p = 0.692$ \\
    \path{has_remove} & 114 & 4.56; $p = 0.002^{**}$ & 2.50; $p = 0.006^{**}$ \\
    \path{has_adjust} & 653 & 1.21; $p = 0.711$ & 0.55; $p = 0.069$ \\

    \midrule
    \rowcolor{LightGray} \multicolumn{4}{l}{\textbf{Refactoring Dimension: Scope}} \\
    \midrule

    \path{has_method} & 312 & 0.22; $p = 0.003^{**}$ & 0.23; $p = 4.59\mathrm{e}{-06}^{***}$ \\
    \path{has_class} & 168 & 1.07; $p = 0.898$ & 0.57; $p = 0.111$ \\
    \path{has_local_variable} & 441 & 1.37; $p = 0.541$ & 1.25; $p = 0.507$ \\

    \bottomrule
  \end{tabularx}
\end{table*}

\begin{tcolorbox}[colframe=black,colback=gray!20]
\textbf{RQ2 Summary.}
The presence of tangled refactorings in agent-generated patches is significantly associated with reduced compilation success, while exhibiting no statistically significant association with issue resolution success. 
At a finer granularity, method-level tangled refactorings are consistently negatively associated with agent effectiveness in issue resolution, whereas removal-oriented tangled refactorings are positively associated with improved effectiveness.
\end{tcolorbox}

\subsection{RQ3: Effectiveness of \tech{} for Improving Issue Resolution}
% ~\tz{to be rewritten @ tian}

\subsubsection{Approach}
Building on the findings from RQ2, which show that refactoring presence (particularly method-level refactorings) is significantly negatively associated with coding agents' effectiveness in issue resolution, we propose a lightweight refactoring-aware approach, \textbf{\tech{}}, as a post-processing module for existing coding agents.
The goal of \tech{} is to assess and selectively transform tangled refactoring operations in generated patches to improve overall patch quality.
Specifically, \tech{} consists of two main components: (1) \textbf{Refactoring Assessment Component}, which evaluates the quality of detected tangled refactoring operations, and (2) \textbf{Patch Refinement Component}, which selectively modifies or removes low-quality tangled refactorings to enhance patch compilability and correctness.

% \begin{itemize}
%     \item \textbf{Necessity}: This dimension determines whether the refactoring is semantically tangled with the bug fix logic and thus required for the correctness of the fix, or whether it constitutes an independent, non-essential structural change that can be separated from the logic of the fix~\cite{Ferreira2025BugPronenessMixed};
%     \item \textbf{Safety}: This dimension assesses whether the refactoring maintains behavioral and structural consistency when tangled with the bug-fixing logic, or whether this introduces inconsistencies such as type errors, broken call sites, or dependency violations.
% \end{itemize}

\textbf{Refactoring Assessment Component.}
For each agent-generated patch, we provide \tech{} with the issue description, the generated patch, and a structured list of detected tangled refactorings (including their type, location, and descriptions produced by RefactoringMiner~\cite{Alikhanifard:TOSEM:2024:RefactoringMiner3.0}). 
Based on this information, \tech{} assesses each refactoring operation.
Specifically, \tech{} employs a base LLM (we used DeepSeek-V3.2 or Gemini-2.5-Pro in RQ3) to evaluate each tangled refactoring along two orthogonal dimensions that capture its necessity and safety within the context of issue resolution:
(1) \textbf{Necessity}, which determines whether the refactoring is semantically intertwined with the logic required to resolve the issue, or instead represents an independent, non-essential structural modification; and
(2) \textbf{Safety}, which evaluates whether the refactoring preserves behavioral and structural consistency when combined with the issue-resolution logic, or introduces inconsistencies such as type errors, broken call sites, or dependency violations.

Based on this assessment, \tech{} assigns one of three actions: 
\path{KEEP}, \path{REMOVE}, or \path{FIX}. 
The decision rules are defined as follows: 
(1) refactoring operations that are both necessary and safe are labeled \path{KEEP}; 
(2) those that are necessary but unsafe are labeled \path{FIX} (accompanied by natural-language refinement suggestions); 
and (3) those deemed unnecessary are labeled \path{REMOVE}, regardless of their safety.

In this research question, we focus on the 791 agent-generated patches that contain at least one tangled refactoring (as reported in Table~\ref{tab:refactoring_quadrants}). 
The assessment stage processes these patches, covering 2,263 tangled refactorings across 72 distinct refactoring types for Gemini-2.5-Pro. 
Four patches are excluded due to exceeding the context window of DeepSeek-V3.2 when combining patch diffs and refactoring context. 
As a result, 787 patches are successfully processed, yielding assessments for 1,812 tangled refactoring instances for DeepSeek-V3.2.

% \footnote{Four instances were excluded because the combined length of the patch diff and refactoring context exceeded the DeepSeek V-3.2's context window. Among the remaining 787 instances, 451 operations in 11 high-volume instances (20--90 operations each) were removed due to output truncation; these instances are retained based on the assessments produced.}

\textbf{Patch Refinement Component.}
Based on the assessment outcomes, we categorize tangled refactorings into three action types: \path{KEEP}, \path{REMOVE}, and \path{FIX}. 
If all tangled refactorings are labeled \path{KEEP}, no modification is performed, and the refined patch remains identical to the original. 
Otherwise, when at least one refactoring is labeled \path{REMOVE} or \path{FIX}, \tech{} invokes an LLM to generate a refined patch by leveraging both the original patch and the structured assessment outputs.
The refinement process is guided by the following objectives: (1) eliminating tangled refactorings labeled \path{REMOVE}; 
(2) applying corrective modifications to tangled refactorings labeled \path{FIX}; 
(3) preserving tangled refactorings labeled \path{KEEP}; 
and (4) maintaining the intended compilability and correctness of the patch for issue resolution.
To evaluate the effectiveness of the patch refinement component, we adopt the evaluation tool provided by Multi-SWE-bench\footnote{\url{https://github.com/multi-swe-bench/multi-swe-bench}} to measure both compilability and correctness of the refined patches.
In addition, the patch refinement component employs DeepSeek-V3.2 or Gemini-2.5-Pro as the base model to ensure computational efficiency and reproducibility under practical resource constraints. 
We set the temperature to 0 to enforce deterministic decoding and consistent outputs.

\begin{table}[t]
\centering
\tabcolsep=2.5mm
\normalsize
\caption{Joint distribution of necessity and safety assessments for tangled refactorings, with the corresponding recommended action.}
\label{tab:necessity_safety}
\resizebox{\textwidth}{!}{
    \begin{tabular}{lll|rrr}
    \toprule
    \textbf{Base LLM} & \textbf{Necessity} & \textbf{Safety} & \textbf{\# Refactorings} & \textbf{\% Refactorings} & \textbf{Action} \\
    \midrule
    \multirow{5}{*}{DeepSeek-V3.2} & Necessary     & Safe & 583 & 32.17\% & \path{KEEP}   \\
    & Necessary     & Unsafe & 396 & 21.85\% & \path{FIX}    \\
    & Unnecessary  & Safe &209 & 11.53\% & \path{REMOVE} \\
   &  Unnecessary  & Unsafe& \textbf{624} & \textbf{34.44\%} & \path{REMOVE} \\
    \cmidrule(lr){2-6}
    & \multicolumn{2}{c|}{\textbf{Total}}& \textbf{1{,}812} & \textbf{100\%} & --- \\
    \midrule
    \multirow{5}{*}{Gemini-2.5-Pro} & Necessary     & Safe & 835 & 35.29\% & \path{KEEP}   \\
    & Necessary     & Unsafe & 154 & 6.51\% & \path{FIX}    \\
    & Unnecessary  & Safe &408 & 17.24\% & \path{REMOVE} \\
   &  Unnecessary  & Unsafe& \textbf{969} & \textbf{40.96\%} & \path{REMOVE} \\
    \cmidrule(lr){2-6}
    & \multicolumn{2}{c|}{\textbf{Total}}& \textbf{2{,}366} & \textbf{100\%} & --- \\
    \bottomrule
    \end{tabular}}
\end{table}

\textbf{Implementation of \tech{}.} 
Specifically, we apply a unified prompt template for \tech{} across all instances. 
The prompts are designed following established best practices in prompt engineering~\cite{prompt_engneering1}.
For both components of \tech{}, we adopt a three-layer prompt design:
(1) \textbf{Global system prompt}, which defines shared priorities across both components (i.e., refactoring assessment and patch refinement), including correctness-first principles, minimal change scope, disciplined use of refactoring, backward compatibility, and preservation of test interfaces;
(2) \textbf{Component-specific system prompt}, which specifies task-specific rules and output constraints. 
For example, in the refactoring assessment component, it enforces the evaluation of each refactoring operation along necessity and safety dimensions, whereas in the patch refinement component, it requires outputs in a unified diff format to ensure successful patch application;
(3) \textbf{Case-specific user prompt}, which provides instance-level inputs.
For the refactoring assessment component, this includes the issue description, relevant code context, the original patch diff, and the detected refactoring list.
For the patch refinement component, it includes the original patch diff, assessment results, and per-refactoring actions (i.e., \path{KEEP}: unchanged; \path{REMOVE}: revert refactoring-only edits; \path{FIX}: apply refinement suggestions), while constraining the output to a single executable unified Git diff.
Complete prompt templates are provided in Appendix~\ref{app:rq3_prompts}.

% \begin{table}[t]
% \centering
% \tabcolsep=2.5mm
% \normalsize
% \caption{Improvement of \tech{} over existing coding agents in terms of compatibility.}
% \label{tab:refinement_compilation}
% \begin{threeparttable}
% \begin{tabular}{lccc|cc}
% \toprule
% \textbf{Agent} & \textbf{\# Patches} & \textbf{Original\tnote{*}} & \textbf{w/ \tech{}\tnote{*}} & \textbf{F2S\tnote{\dag} \, ($\uparrow$)} & \textbf{S2F\tnote{\dag} \, ($\downarrow$)} \\
% \midrule
% SWE-agent  & 250 & 50 (20.00\%) & 76 (30.40\%) & 30 (12.00\%) & 4 (1.60\%) \\
% Agentless  & 150 & 27 (18.00\%) & 40 (26.67\%) & 15 (10.00\%) & 2 (1.33\%) \\
% OpenHands  &  76 & 17 (22.37\%) & 25 (32.89\%) &  8 (10.53\%) & 0 (0\%) \\
% \midrule
% \textbf{Total} & \textbf{476} & \textbf{94 (19.75\%)} & \textbf{141 (29.62\%)} & \textbf{53 (11.13\%)} & \textbf{6 (1.26\%)} \\
% \bottomrule
% \end{tabular}
% \begin{tablenotes}[flushleft]
%     % \scriptsize
%     \footnotesize
%     \setlength{\itemsep}{0pt}
%     \item[*] \textbf{Original} and \textbf{w/ \tech{}} represent the original coding agent and the combination of the coding agent and \tech{}, respectively.
%     \item[\dag] \textbf{F2S} and \textbf{S2F} denote compatibility transitioning from failure to success and from success to failure, respectively.
% \end{tablenotes}
% \end{threeparttable}
% \end{table}

\begin{table}[t]
\centering
\tabcolsep=1.9mm
% \normalsize
\small
\caption{Improvement of \tech{} over existing coding agents in terms of compatibility.}
\label{tab:refinement_compilation}
\begin{threeparttable}
\begin{tabular}{llc|cc|cc}
\toprule
\multirow{2}{*}{\textbf{Base LLM}} & \multirow{2}{*}{\textbf{Agent}} & \textbf{\# Valid} & \multicolumn{2}{c|}{\textbf{Compilable Patch ($\uparrow$})} & \multirow{2}{*}{\textbf{F2S\tnote{\dag} \, ($\uparrow$)}} & \multirow{2}{*}{\textbf{S2F\tnote{\dag} \, ($\downarrow$)}} \\ \cmidrule(lr){4-5}
% \textbf{Base LLM} & \textbf{Agent} & \textbf{\# Patches} & \textbf{Original\tnote{*}} & \textbf{w/ \tech{}\tnote{*}} & \textbf{F2S\tnote{\dag} \, ($\uparrow$)} & \textbf{S2F\tnote{\dag} \, ($\downarrow$)} \\
& & \textbf{Patches} & \textbf{Original\tnote{*}} & \textbf{w/ \tech{}\tnote{*}} & & \\
\midrule
\multirow{4}{*}{DeepSeek-V3.2} & SWE-agent & 250 & 50 (20.00\%) & 76 (30.40\%) & 30 (12.00\%) & 4 (1.60\%) \\
& Agentless & 150 & 27 (18.00\%) & 40 (26.67\%) & 15 (10.00\%) & 2 (1.33\%) \\
& OpenHands & 76 & 17 (22.37\%) & 25 (32.89\%) &  8 (10.53\%) & 0 (0\%) \\ \cmidrule(lr){2-7}
& \textbf{Total} & \textbf{476}& \textbf{94 (19.75\%)} & \textbf{141 (29.62\%)} & \textbf{53 (11.13\%)} & \textbf{6 (1.26\%)} \\
\midrule
\multirow{4}{*}{Gemini-2.5-Pro} & SWE-agent & 221 & 37 (16.74\%) & 107 (48.42\%) & 75 (33.94\%) & 5 (2.26\%) \\
& Agentless & 99 & 24 (24.24\%) & 44 (44.44\%) & 21 (21.21\%) & 1 (1.01\%) \\
& OpenHands & 50 & 9 (18.00\%) & 23 (46.00\%) & 14  (28.00\%) & 0 (0\%) \\ \cmidrule(lr){2-7}
& \textbf{Total} & \textbf{370} & \textbf{70 (18.92\%)} & \textbf{174 (47.03\%)} & \textbf{110 (29.73\%)} & \textbf{6 (1.62\%)} \\
\bottomrule
\end{tabular}
\begin{tablenotes}[flushleft]
    % \scriptsize
    \footnotesize
    \setlength{\itemsep}{0pt}
    \item[*] \textbf{Original} and \textbf{w/ \tech{}} represent the original coding agent and the combination of the coding agent and \tech{}, respectively.
    \item[\dag] \textbf{F2S} and \textbf{S2F} denote compatibility transitioning from failure to success and from success to failure, respectively.
\end{tablenotes}
\end{threeparttable}
\end{table}

\subsubsection{Results}
Table~\ref{tab:necessity_safety} presents the joint distribution of necessity and safety assessments across 1,812 (for DeepSeek-V3.2) and 2,366 (for Gemini-2.5-Pro) evaluated tangled refactorings, while Table~\ref{tab:refinement_compilation} reports the performance improvements of \tech{} over baseline coding agents in terms of compilability and correctness.

\textbf{Refactoring Assessment.}
As shown in Table~\ref{tab:necessity_safety}, the largest category of tangled refactorings is \textit{unnecessary-and-unsafe} (34.44\% or 40.96\%), whereas \textit{necessary-and-safe} refactorings account for only 32.17\% or 35.29\%. 
This indicates a substantial proportion of tangled refactorings that are both irrelevant to issue resolution and incorrectly implemented, which may adversely affect agent effectiveness in issue resolution. 
For DeepSeek-V3.2, \textit{necessary-but-unsafe} refactorings (21.85\%) represent another major source of risk, as these correspond to structurally required changes that are incompletely or incorrectly realized (e.g., missing call-site updates or type inconsistencies); for Gemini-2.5-Pro, this type only accounts for 6.51\% of refactorings.
By contrast, \textit{unnecessary-but-safe} refactorings (11.53\% for DeepSeek-V3.2) are correctly implemented but introduce superfluous modifications that do not contribute to resolving the issue; for Gemini-2.5-Pro, this type increases to 17.24\% of refactorings.

At the patch level, among the 787 for DeepSeek-V3.2 assessed patches, 561 (71.28\%) or contain at least one unnecessary or unsafe tangled refactoring and therefore require refinement, while only 226 (28.72\%)  consist exclusively of necessary and safely implemented refactorings. In comparison, Gemini-2.5-Pro is more conservative, yielding a nearly even split between patches marked as needing refinement and those kept as original, with 399 (50.44\%) and 392 (49.56\%) patches, respectively.
This distribution suggests that substantial portions of agent-generated patches include tangled refactorings that are either unnecessary or potentially harmful. 
Notably, this observation aligns with the negative association between refactoring presence and compilability identified in RQ2 (Section~\ref{sec: 2}).

\textbf{Patch Refinement.}
The patch refinement component also employs either DeepSeek-V3.2 or Gemini-2.5-Pro as the base model to balance computational efficiency and reproducibility under practical resource constraints. 
Specifically, among the 561 patches for DeepSeek-V3.2 and 399 patches for Gemini-2.5-Pro identified as requiring refinement, we exclude patches that already successfully resolve the target issue, as well as patches that cannot be processed due to LLM context window limitations. 
As a result, 476 patches are refined using DeepSeek-V3.2, while 370 patches are refined using Gemini-2.5-Pro for subsequent evaluation.
% Among the 561 patches requiring refinement, we exclude 57 patches that already successfully resolve the issue and 28 patches that cannot be processed due to LLM context window limitations, resulting in 476 refined patches for evaluation. 
We denote transitions in compilability as \textbf{\textit{F2S}} (failure-to-success) and \textbf{\textit{S2F}} (success-to-failure).

Overall, \tech{} achieves substantial improvements in patch compilability. 
Specifically, 53 (11.13\%) and 110 (29.73\%) patches transition from failure to success after refinement (\textit{F2S}) when using DeepSeek-V3.2 and Gemini-2.5-Pro, respectively, while only 6 patches regress from success to failure (\textit{S2F}) under each setting (1.26\% for DeepSeek-V3.2 and 1.62\% for Gemini-2.5-Pro). 
Using DeepSeek-V3.2 as the refinement model, the compilation success rate increases from 19.75\% (94/476) to 29.62\% (141/476), representing a relative improvement of 50\%. 
When using Gemini-2.5-Pro, the compilation success rate further increases from 18.92\% (70/370) to 47.03\% (174/370), corresponding to a relative improvement of 149\%. 
Notably, these improvements are consistently observed across all three coding agents.
These results demonstrate that our refactoring-aware refinement approach \tech{} can effectively mitigate structural issues that hinder successful patch compilability.
Moreover, the substantially larger gains achieved with the more advanced Gemini-2.5-Pro suggest that the effectiveness of \tech{} further improves as the capability of the underlying base model increases.

In terms of correctness, \tech{} also demonstrates promising improvements. 
Specifically, 6 (1.26\%) and 16 (4.32\%) previously unresolved patches are successfully refined to pass all test cases when using DeepSeek-V3.2 and Gemini-2.5-Pro, respectively.
These results suggest that correcting problematic tangled refactorings can, in certain instances, directly enable successful issue resolution. 
Although the improvements in correctness are more modest than those observed for compilability, the results nonetheless highlight the potential of refactoring-aware refinement for improving functional correctness.
Furthermore, the larger gains achieved with Gemini-2.5-Pro compared to DeepSeek-V3.2 indicate that the effectiveness of \tech{} benefits from stronger underlying base models.
Future work may further enhance these gains by incorporating more advanced LLMs and more sophisticated refinement strategies.

\begin{tcolorbox}[colframe=black,colback=gray!20]
\textbf{RQ3 Summary.}
% Refactoring-aware assessment shows that unnecessary or unsafe refactorings are common in agent-generated patches: 561/787 patches (71.3\%) include at least one operation labeled \path{REMOVE} or \path{FIX}. Applying targeted refinement to 476 eligible patches improves compilation success from 19.7\% (94/476) to 29.6\% (141/476), with 53 improvements versus 6 regressions (about 9:1). Beyond compilation gains, refinement also turns 6 previously unresolved patches into fully passing solutions, indicating that correcting refactoring-related issues can directly recover patch correctness.
Refactoring-aware assessment reveals that unnecessary or unsafe refactorings are prevalent in agent-generated patches, affecting 50.44--71.28\% of the analyzed instances.
Our proposed \tech{} substantially improves patch compilability, increasing the average compilation success rate from 19.34\% to 38.33\%.
Beyond these gains, \tech{} also enables an average of 2.79\% previously unresolved patches to pass all test cases, highlighting its potential for improving functional correctness.
\end{tcolorbox}

% \textbf{Regression analysis.}
% The 6 compilation regressions (instances where the refined patch fails to compile but the original succeeded) all occurred in SWE-agent (4 instances) and AgentLess (2 instances) patches, with none in OpenHands.
% These cases arise when the refinement excises or modifies code that, despite being assessed as problematic refactoring, was structurally load-bearing within the patch.
% This highlights an inherent limitation of the two-stage pipeline: the assessment may occasionally misjudge the necessity of an operation, or the refinement may disrupt inter-dependent code structures during modification.
% Nevertheless, the strongly favorable 9:1 improvement-to-regression ratio confirms that the pipeline is robustly net-positive in practice.

\section{Discussion}
\subsection{Case Study for RQ2: Impact of Refactoring Practices on Patch Correctness}
\label{dis:RQ2}

In RQ2, we find that method-level tangled refactorings are significantly negatively associated with both compilation success (OR of 0.24 with $p < 0.001$) and successful issue resolution (OR of 0.22 with $p = 0.002$).
To illustrate how such method-level tangled refactorings can introduce compilation failures in practice, we present a representative case study based on the instance \path{fasterxml__jackson-databind-3860}. 
According to the corresponding issue report,\footnote{\url{https://github.com/FasterXML/jackson-databind/issues/3814}} the task requires enhancing the abstract class \path{StdNodeBasedDeserializer<T>} in the Jackson-databind library to support the \path{ObjectMapper #readerForUpdating} workflow.
Among the 34 valid agent-generated patches for this issue, 24 compile successfully, and 22 ultimately pass all tests and correctly resolve the issue.

% \tao{also resolve? we want to also show this}
% % kind of refactoring risk to issue resolution successfully
% \tao{double check these terms work in Java}
% \tao{line number to patch number}
% 24 out of 34 valid agent-generated patches are complied successfully and 22 of them correctly solve the issue. 
For the correct patches, the 22 successful agent-generated patches and the human-written golden patch follow a highly consistent repair strategy. 
Specifically, they (i) introduce a new method, \texttt{convert(JsonNode root, DeserializationContext ctxt, T intoValue)}, with a default implementation that forwards calls to the existing two-parameter \path{convert} (Lines~4--8); and (ii) add a new overridden method, \texttt{deserialize(JsonParser jp, DeserializationContext ctxt, T intoValue)}, which invokes the three-parameter \path{convert} method while preserving the original two-parameter method signature (Lines~16--21). 
Notably, this repair strategy does not involve refactoring operations detectable by RefactoringMiner~\cite{Alikhanifard:TOSEM:2024:RefactoringMiner3.0}.
A simplified view of the correct patch is shown below:

\begin{lstlisting}[language=diff]
@@ StdNodeBasedDeserializer.java
   public abstract T convert(JsonNode root, DeserializationContext ctxt) throws IOException;

+  public T convert(JsonNode root, DeserializationContext ctxt, T intoValue)
+      throws IOException {
+    ctxt.handleBadMerge(this);
+    return convert(root, ctxt);
+  }
   
   @Override
   public T deserialize(JsonParser jp, DeserializationContext ctxt) throws IOException {
     JsonNode n = _treeDeserializer.deserialize(jp, ctxt);
     return convert(n, ctxt);
   }

+  @Override
+  public T deserialize(JsonParser jp, DeserializationContext ctxt, T intoValue)
+      throws IOException {
+    JsonNode n = _treeDeserializer.deserialize(jp, ctxt);
+    return convert(n, ctxt, intoValue);
+  }
\end{lstlisting}

For the incorrect agent-generated patches, three patches containing method-level tangled refactorings failed to compile and consequently failed to resolve the issue.
As an illustrative example, the patch generated by Agentless combined with Qwen2.5-72B introduces two method-level tangled refactoring operations: 
(i) \emph{Add Parameter}, which adds the parameter \path{intoValue} to the existing \path{deserialize} method (Lines~3--6); 
and (ii) \emph{Extract Method}, which extracts the conversion logic into a new \path{convert} method (Lines~8--15), instead of introducing the required overridden method.
A simplified view of the incorrect patch is shown below:

\begin{lstlisting}[language=diff]
@@ StdNodeBasedDeserializer.java
   @Override
-  public T deserialize(JsonParser jp, DeserializationContext ctxt) 
-      throws IOException {
+  public T deserialize(JsonParser jp, DeserializationContext ctxt, T intoValue)
+      throws IOException {
     JsonNode n = _treeDeserializer.deserialize(jp, ctxt);
-    return convert(n, ctxt);
+    return convert(n, ctxt, intoValue);
+  }
+
+  public T convert(JsonNode root, DeserializationContext ctxt, T intoValue)
+      throws IOException {
+    ctxt.handleBadMerge(this);
+    return convert(root, ctxt);
   }
\end{lstlisting}

% \tao{make sure words are used commonly, you can directly find its usage in google scholar or google with keywords: "class hierarchy contract"} \tao{contract -> annotation?}

Notably, both \emph{Add Parameter} and \emph{Extract Method} are categorized as method-level refactoring operations under the Refactoring Scope dimension (Table~\ref{tab:rq2_variables}), and therefore contribute to the independent variable of \path{has_method}.
Specifically, the \emph{Add Parameter} refactoring modifies the original two-parameter \path{deserialize} method signature into a three-parameter variant, while the \emph{Extract Method} operation relocates the conversion logic into a new \path{convert} method.
However, because \path{StdNodeBasedDeserializer<T>} extends \path{StdDeserializer<T>}, the original two-parameter method \path{deserialize(JsonParser, DeserializationContext)} must be overridden to satisfy the inheritance hierarchy.
The combination of these two method-level refactorings alters the required method signature instead of introducing a valid overloaded override, thereby violating the \path{@Override} contract enforced by the Java type system and causing a compilation failure.
Consequently, the generated patch triggers the compiler error indicating that it ``does not override abstract method \path{deserialize(JsonParser, DeserializationContext)} in \path{JsonDeseriablizer}.''
% \tao{specific error message}

% \tao{what about issue resolution}
% \tao{make sure commonly use term}
% \tao{same above}
% \tao{use the format consistently across paper for refactoring type emuf}
This case study provides a concrete explanation for why \path{has_method} is strongly negatively associated with compilation success (OR\,=\,0.24, $p = 4.59\mathrm{e}{-06}$) in RQ2.
Specifically, method-level refactoring operations such as \emph{Add Parameter} and \emph{Extract Method} modify existing method structures and signatures rather than introducing purely additive functionality. 
In nominally typed object-oriented languages such as Java, method overriding requires exact signature matching. Consequently, structural modifications to inherited method signatures may violate the \path{@Override} contract defined by the superclass, resulting in compilation failures and, ultimately, unsuccessful issue resolution.
% Among the 33 coding agents, all three coding agents whose patches contained method-level refactoring failed compilation, while 24 of the 33 coding agents without method-level refactoring compiled successfully. Among 24 of them, 22 patches successfully resolve corresponding issue.
% \tao{also discuss more about issue resolution.}
% \tao{number is not consistent 33 or 30}
% \tao{term consistently coding agent or agent}

\subsection{Case Study for RQ3: Refactoring-Aware Patch Refinement}

% \tao{ structural issues? make sure term consistent with RQ3}
% \tao{avoid using ---}

In RQ3, we find that refactoring-aware assessment identifies at least one unnecessary or unsafe refactoring operation in 50.44--71.28\% of the analyzed patches. 
Subsequent refinement using DeepSeek-V3.2 improves the compilation success rate from 19.75\% to 29.62\% and enables 1.26\% of previously unresolved instances to be successfully resolved.
Similarly, refinement based on Gemini-2.5-Pro increases the compilation success rate from 18.92\% to 47.03\% and enables 4.32\% of previously unresolved instances to be successfully resolved.
To illustrate how our two-stage pipeline, refactoring assessment followed by patch refinement, can recover functional correctness, we present a representative case study based on the instance \path{google__gson-1555}. 
According to the corresponding issue report,\footnote{\url{https://github.com/google/gson/issues/1553}} the \path{@JsonAdapter} annotation in the Gson library defines a \path{nullSafe} parameter; however, this parameter is ignored when the adapter is implemented as a \path{JsonSerializer} or \path{JsonDeserializer}.
Specifically, the internal class \path{TreeTypeAdapter} unconditionally returns \path{null} for JSON null inputs and serializes \path{null} for null objects. 
The issue report further notes that \path{TreeTypeAdapter} behaves as permanently null-safe and therefore proposes introducing an explicit \path{nullSafe} parameter, which is precisely the repair strategy adopted in the human-written golden patch.

The human-written golden patch introduces a new \path{nullSafe} field and propagates it as a constructor parameter to \path{TreeTypeAdapter} (Lines~2--8), while also adding a backward-compatible constructor overload.
In addition, the patch updates the \path{read()} and \path{write()} methods to conditionally handle null values based on the \path{nullSafe} flag (Lines~10--13), and modifies the external call site in \path{JsonAdapterAnnotationTypeAdapterFactory} accordingly (Lines~28--30). 
A simplified view of the core repair logic is shown below:

% \tao{@@ to be consistent}
\begin{lstlisting}[language=diff]
@@ TreeTypeAdapter.java
+  private final boolean nullSafe;
   public TreeTypeAdapter(JsonSerializer<T> serializer, JsonDeserializer<T> deserializer,
-      Gson gson, TypeToken<T> typeToken, TypeAdapterFactory skipPast) {
+      Gson gson, TypeToken<T> typeToken, TypeAdapterFactory skipPast, boolean nullSafe) {
     ...
+    this.nullSafe = nullSafe;
+  }
+ 
+  public TreeTypeAdapter(JsonSerializer<T> serializer, JsonDeserializer<T> deserializer,
+      Gson gson, TypeToken<T> typeToken, TypeAdapterFactory skipPast) {
+    this(serializer, deserializer, gson, typeToken, skipPast, true);
+  }
   ...
-    if (value.isJsonNull()) {
+    if (nullSafe && value.isJsonNull()) {
       return null;
     }
   ...
-    if (value == null) {
+    if (nullSafe && value == null) {
       out.nullValue();
       return;
     }
   ...

@@ JsonAdapterAnnotationTypeAdapterFactory.java
-      typeAdapter = new TreeTypeAdapter(serializer, deserializer, gson, type, null);
+      typeAdapter = new TreeTypeAdapter(serializer, deserializer, gson, type, null, 
+          annotation.nullSafe());
\end{lstlisting}

% \tao{use line numbers}
% \tao{I think here is three patches one golden patch; one agent patch; one our refined patch; refined patch is missing}
The agent-generated patch (produced by SWE-agent with GPT-4o) correctly captures the essential repair logic by introducing the \path{nullSafe} parameter, implementing conditional null handling, and updating the corresponding call sites.
However, the patch also introduces two refactoring operations detected by RefactoringMiner.
The first is an \emph{Add Parameter} refactoring, which modifies the existing \path{TreeTypeAdapter} constructor by extending its signature from five to six parameters through the addition of \path{nullSafe}, instead of introducing a backward-compatible constructor overload as in the golden patch (Lines~21--26).
The second operation is a \emph{Move Class} refactoring, where the agent-generated patch erroneously replaces the copyright header and \path{package} declaration at the beginning of \path{TreeTypeAdapter.java} with the contents of the unrelated test class \path{NullSafeTest} (Lines~2--15).
A simplified view of the original agent-generated patch is shown below:

\begin{lstlisting}[language=diff]
@@ TreeTypeAdapter.java
-  /*
-   * Copyright (C) 2011 Google Inc.
-   * ... (license header)
-   */
-  package com.google.gson.internal.bind;
+  import com.google.gson.*;
+  ...
+  class NullSafeTest {  // [Refactoring: Move Class]
+    @JsonAdapter(value = PersonAdapter.class,
+        nullSafe = false)
+    static class PersonAdapter extends TypeAdapter<Person>{
+    ...
    }
+  }
   import com.google.gson.Gson;  // resumes
   ...

+  private final boolean nullSafe;
   public TreeTypeAdapter(...,
-      TypeAdapterFactory skipPast) {
+      TypeAdapterFactory skipPast,
+      boolean nullSafe) {  // [Refactoring: Add Parameter]
     ...
+    this.nullSafe = nullSafe;
   }
   ...
-    if (value.isJsonNull()) {
+    if (value.isJsonNull() && nullSafe) {
       return null;
     }
   ...
-    if (value == null) {
+    if (value == null && nullSafe) {
       out.nullValue(); return;
     }
   ...
+  // Internal call site (SingleTypeFactory.create):
+  return matches
+      ? new TreeTypeAdapter<T>((JsonSerializer<T>) serializer,
-              (JsonDeserializer<T>) deserializer, gson, type, this)
+              (JsonDeserializer<T>) deserializer, gson, type, this, true)
+      : null;

@@ JsonAdapterAnnotationTypeAdapterFactory.java
-      typeAdapter = new TreeTypeAdapter(serializer, deserializer, gson, type, null);
+      typeAdapter = new TreeTypeAdapter(serializer, deserializer, gson, type, null,
+          annotation.nullSafe());
\end{lstlisting}

% \tao{error message}
% \tao{for all patches use // inline to marker this line is detected refactor, for example TypeAdapterFactory skipPast , //detected as xx refactoring, if multiple lines show multiple // comments}
The original agent-generated patch fails compilation and subsequently fails to resolve the issue.
Specifically, the erroneous \emph{Move Class} operation removes the required \path{package} declaration and introduces a duplicate class definition, resulting in naming conflicts and structural corruption of the source file.

During the refactoring assessment, \tech{} first evaluates each detected refactoring along the necessity and safety dimensions.
The \emph{Move Class} operation is classified as both \emph{unnecessary} and \emph{unsafe}, as it constitutes an irrelevant and structurally invalid modification that replaces the original file header with unrelated test code.
Consequently, it is assigned the refined action \path{REMOVE}.
In contrast, the \emph{Add Parameter} operation is assessed as \emph{required} but \emph{unsafe} (introducing the \path{nullSafe} parameter is required to implement the intended functionality, but the surrounding file corruption prevents the patch from remaining structurally valid).
Therefore, this refactoring is assigned the refined action \path{FIX}, accompanied by a refinement suggestion to restore the original file structure. 
Based on these assessment results, the patch is marked for refinement.
% \tao{since both remove and fix, we then refine right, make sure this is consistent with  RQ3}

During patch refinement, \tech{} receives both the original agent-generated patch diff and the structured assessment output.
Guided by the \path{REMOVE} instruction, the refinement module eliminates the unrelated test code embedded in the file header and restores the original copyright notice and \path{package} declaration.
For the \path{FIX}-labeled \emph{Add Parameter}, the refinement determines that the parameter addition itself is functionally correct, including the constructor update, field assignment, conditional logic in \path{read()} and \path{write()}, and external call-site modifications; therefore, it preserves them unchanged.
As a result, the refinement process only removes the structurally invalid modifications while retaining the semantically necessary repair logic.
The refined patch subsequently compiles successfully and passes all test cases in the Gson test suite, thereby fully resolving the issue.
% \tao{the missing refined patches}

This case study demonstrates the end-to-end effectiveness of our refactoring-aware refinement pipeline.
Although the coding agent correctly implemented the core repair logic, including the introduction of the \path{nullSafe} parameter and the associated conditional null handling, the generated patch also contained an unnecessary and unsafe refactoring operation (\emph{Move Class}) that corrupted the source file structure and ultimately caused compilation failure.
The assessment stage successfully identified the \emph{Move Class} operation as requiring \path{REMOVE}, while flagging the \emph{Add Parameter} operation for further refinement.
Guided by these assessment results, the refinement stage removed the structurally corrupted file header while preserving the semantically necessary feature implementation intact.
More broadly, this example illustrates a common failure pattern observed in agent-generated patches, i.e., a functionally correct repair becomes invalid due to unnecessary structural modifications introduced during patch generation. 
This pattern is representative of the compilation recoveries and successful issue resolutions achieved by our pipeline in RQ3.

\subsection{Implications}
% \tao{do word-to-word check}
% Our findings have practical implications for three key stakeholders: researchers, coding agent users, and coding agent framework developers. 
Because coding agents are jointly shaped by both the agent framework and the underlying LLM, our findings carry practical implications for three groups of stakeholders: \textbf{coding agent framework designers}, \textbf{LLM providers}, and \textbf{practitioners who adopt coding agents for software maintenance tasks}.

% \tao{encoded?} 
% \tao{make sure terms commonly used}
% \tao{same above}
% \tao{distinguish coding agent, agent framework, base LLMs}

\textbf{Implications for Coding Agent Framework Designers.}
Our results show that agent frameworks with higher autonomy (e.g., SWE-agent) tend to introduce substantially more tangled refactorings, whereas structured pipeline-based frameworks (e.g., Agentless) constrain generated patches toward more localized and targeted modifications. 
These findings suggest that agent-framework architecture should be treated as a key factor influencing the correctness and structural quality of AI-generated patches.
Moreover, our preliminary experiments on the two-stage refinement pipeline demonstrate that identifying and correcting problematic refactoring operations after patch generation can effectively improve patch quality. 
This observation indicates that lightweight post-processing mechanisms may serve as a practical and model-agnostic strategy for improving existing coding agents.
To better balance agent autonomy with structural safety, framework designers may consider incorporating mechanisms such as: (1) restricting large-scale refactorings unless explicitly required by the issue context; (2) validating method-signature modifications against inheritance hierarchies and interface contracts; and (3) separating refactoring-related transformations from bug-fix generation into independently verifiable stages.

\textbf{Implications for LLM Providers.} 
Our findings suggest that coding agents built upon specific LLMs tend to reproduce refactoring patterns implicitly learned from training data, yet often lack the contextual reasoning required to determine whether such refactorings are necessary and safe in a given maintenance scenario. 
As illustrated in RQ2 (Section~\ref{dis:RQ2}), coding agents may generate method-level refactorings (e.g., \emph{Add Parameter}) that alter inherited method signatures and unintentionally violate overriding constraints in nominally typed object-oriented languages such as Java.
Importantly, these failure patterns resemble those frequently observed in real-world software commits, where refactoring operations are commonly tangled with bug-fixing logic. 
This suggests that current LLMs may implicitly learn undesirable maintenance patterns from large-scale software repositories. 
Therefore, we encourage LLM providers to further investigate how training-data construction and curation strategies (e.g., untangling refactoring changes from bug-fix commits) may help mitigate these learned biases in base LLMs.

\textbf{Implications for Practitioners Using Coding Agents.}
For practitioners applying coding agents to software issue resolution, our results indicate that generated patches frequently contain unnecessary or unsafe refactoring operations. 
Unlike intentional human-driven refactorings, these implicit structural modifications are often introduced without explicit design rationale and may lead to compilation failures or incorrect fixes.
In practice, users should carefully examine whether structural modifications (particularly method-level refactorings and signature-changing operations) are genuinely required to resolve the reported issue. 
A practical strategy is to maintain a clear separation between bug-fixing logic and refactoring activities, introducing refactoring only after the correctness of the functional fix has been validated. 
Such separation of concerns can improve patch reliability and simplify debugging, validation, and review.
From the perspective of \textbf{code review}, prior studies have shown that reviewer effort increases when commits mix refactoring with functional changes~\cite{Bacchelli2015untangling}. 
Our findings suggest that this challenge also extends to agent-generated patches. 
Accordingly, reviewers should explicitly verify: (1) whether method-level or signature-changing refactorings are present, and (2) whether these modifications are truly necessary for issue resolution. 
Incorporating refactoring-awareness into both manual and automated review workflows (e.g., flagging method-level refactorings for additional verification) may help reduce the risk of accepting structurally invalid or functionally incorrect patches.

% \section{Threats to Validity}
% \textbf{Construct Validity:} 111/128 resolution rate for golden patch (Multi-SWE-bench's construct)
% \textbf{Internal Validity:} goodness-of-fit , variable choose, refactoringMiner-related. RQ3 : need sampling and mannually check. 
% \textbf{External Validity:} Why only Java. one language for one tool.Python-only refactoring tool cite. refactoring type the more the better. only multi-swe-bench, only java, not general. 

\section{Threats to Validity}

\textbf{Construct Validity.}
% \tao{how this affect RQ2, only RQ3 apply evaluate tools right}. \tao{we should mention how we mitigate, e.g.: To mitigate it, we submit several issue reports to the evaluation tools ...}
% \tao{this is not true right, is not 111 are resolved, is 111 can be evlauted using their tools}
We follow the official Multi-SWE-bench evaluation framework, where a patch is correct if it passes all predefined tests. However, in the Java subset, 17 of 128 human-written golden patches fail the benchmark's own evaluation suite. 
This indicates potential incompleteness or inconsistencies in the benchmark tests, which may bias the issue-resolution variable. 
If the test suite does not fully reflect true semantic correctness, the dependent variable in RQ2 may be noisy, reducing the reliability of the regression analysis.
To mitigate it, we manually examined the failing golden instances and reported evaluation inconsistencies to the official evaluation repository.\footnote{\url{https://github.com/multi-swe-bench/multi-swe-bench/issues/36}} 
\footnote{\url{https://github.com/multi-swe-bench/multi-swe-bench/issues/57}} 
% Although this effort does not eliminate the inherent limitations of the benchmark, it improves transparency regarding the evaluation process and clarifies the observed inconsistencies.
Additionally, we use RefactoringMiner~\cite{Alikhanifard:TOSEM:2024:RefactoringMiner3.0} to identify refactorings.
Although it is one of the most accurate and comprehensive tools for Java refactorings, it may still produce false positives or miss certain refactorings, potentially affecting the precision of our refactoring-related variables.

\textbf{Internal Validity.}
% \tao{this should be range right, we had four models}
In RQ2, our logistic regression models achieve McFadden pseudo-$R^2$ values ranging from 0.19 (compilability) to 0.32 (issue resolution).
While generally acceptable for empirical software engineering studies~\cite{hosmer2013applied}, unobserved confounding factors, such as project-specific coding conventions, test-suite rigor, or fix complexity, may still influence the observed relationships between refactoring practices and patch quality.
To mitigate this threat, our models incorporate multiple control variables derived from prior work~\cite{MAGIS, zan2025multiswebench}.
% Although our variable design draws on prior work~\cite{MAGIS, zan2025multiswebench}, it is not comprehensive and could be expanded to capture additional aspects of patch complexity, such as the number of modified AST nodes.
For RQ3, the refactoring assessment stage relies on LLM-based judgments to determine whether refactoring is necessary or unsafe.
Because tangled refactorings are often intertwined with bug-fixing logic, such judgments can be inherently subjective and context-dependent, even for experienced developers.
As a result, we do not independently validate the LLM assessments through large-scale human annotation.
To reduce this threat, we use structured prompting with explicit evaluation criteria and analyze multiple LLMs to reduce reliance on a single model's preference.
% \wang{check}\tao{correct to me, or change ``analyze'' to ``use'' }
% Future work will incorporate structured manual validation on sampled cases, using LLMs as assistants rather than primary evaluators.
% We also intend to investigate post-validation mechanisms that independently verify refactoring assessments and quantify their impact on downstream patch correctness.
% Moreover, RQ3 evaluates the refinement pipeline using only two LLMs (DeepSeek-V3.2 and Gemini-2.5-Pro), which may introduce model-specific biases. 
% Future work will extend the evaluation to a broader range of foundation models.

\textbf{External Validity.}
Our study focuses exclusively on Java projects, as RefactoringMiner provides comprehensive support for Java refactoring detection, covering more than 100 refactoring types.
Consequently, our findings may not generalize to other programming languages, particularly dynamically typed languages, where refactoring patterns and associated risks may differ substantially.
In addition, our analysis is based on a single benchmark (Multi-SWE-bench), three coding agent frameworks, and twelve foundation models.
Although this setup provides meaningful diversity for empirical evaluation, the results may not fully generalize to other benchmarks, agent architectures, or real-world deployment settings.
Future work should extend the study to broader programming ecosystems and more diverse coding-agent configurations.
% Finally, all evaluated projects are open-source Java repositories, so how coding agents behave in industrial or enterprise-scale software systems (with stricter development standards, more complex dependencies, and proprietary engineering workflows) remains an open question.

\section{Conclusion}

This paper presents the first empirical study of tangled refactorings in agent-based software issue resolution.
Our analysis shows that coding agents perform tangled refactorings less frequently and less intensively than human developers, but exhibit a broader and often riskier range of refactoring behaviors.
We further find that tangled refactorings are significantly associated with reduced patch compilability, while providing no significant benefit to issue resolution success.
Motivated by these findings, we propose \tech{}, a refactoring-aware refinement approach that evaluates the necessity and safety of detected refactorings and selectively removes or repairs problematic ones.
Experimental results show that \tech{} consistently improves both patch compilability and functional correctness, highlighting the potential of refactoring-aware post-generation refinement for improving the reliability of coding agents in software maintenance.
% In future work, we plan to extend this line of research beyond refactoring to other forms of tangled non-functional changes, such as documentation modifications, comment edits, and code-style-only transformations.

\begin{acks}
We gratefully acknowledge the financial support of: (1) JSPS for the KAKENHI grants (25K22845, 26H02500, and 26K21198); (2) Japan Science and Technology Agency (JST) as part of Adopting Sustainable Partnerships for Innovative Research Ecosystem (ASPIRE), Grant Number JPMJAP2415, (3) the Kayamori Foundation of Informational Science Advancement for supporting Tao Xiao, and (4) the Inamori Research Institute for Science for supporting Yasutaka Kamei via the InaRIS Fellowship.
\end{acks}

% future work

% \section*{Data Availability}
%%
%% The acknowledgments section is defined using the "acks" environment
%% (and NOT an unnumbered section). This ensures the proper
%% identification of the section in the article metadata, and the
%% consistent spelling of the heading.
% \begin{acks}
% To Robert, for the bagels and explaining CMYK and color spaces.
% \end{acks}

%%
%% The next two lines define the bibliography style to be used, and
%% the bibliography file.
\bibliographystyle{acm/ACM-Reference-Format}
\bibliography{main}

% \ifincludeappendix

\newpage
\appendix
\section{Full Prompt Templates for \tech{}}\label{app:rq3_prompts}

This appendix reports the full prompts used in \tech{}.
We keep the original wording and placeholder fields for reproducibility.

\subsection{Shared Global System Prompt}

\begin{lstlisting}
You are a senior Java software engineering expert specializing in bug-fix review and refactoring optimization. Your goal is to ensure patches are correct, minimal, and contain only necessary changes.

Key principles:

1. Correctness first: the fix must resolve the reported issue
2. Minimize change scope: avoid unnecessary modifications
3. Refactoring discipline: keep only refactorings that directly support the fix or are tightly coupled to it
4. Maintain backward compatibility: do not introduce breaking changes to public APIs unless absolutely necessary
5. Preserve test interfaces: avoid changes that break existing test contracts

Note: Output format requirements are specified in the phase-specific system prompts.
\end{lstlisting}

\subsection{Component 1: Refactoring Assessment}

\textbf{1. Component-specific system prompt template:}

\begin{lstlisting}
You are a senior Java engineer specializing in patch analysis and refactoring evaluation. Your goal is to assess each refactoring detected in a bug-fix patch and recommend appropriate actions.

Evaluation Dimensions:

1. **Necessity** - Is this refactoring necessary for the fix to work?
   - NECESSARY: The fix cannot work without this refactoring (e.g., must add parameter to pass critical data)
   - UNNECESSARY: The refactoring is cosmetic cleanup unrelated to the fix

2. **Safety** - Is the refactoring implementation correct?
   - SAFE: The refactoring is correctly implemented with no issues
     * All API changes are consistently applied across all call sites
     * Type signatures are compatible or properly updated
     * Behavior is preserved (same logic, proper wiring, equivalent control flow)
   
   - UNSAFE: The refactoring has implementation problems (specify what)
     * Missing updates at some call sites
     * Type incompatibilities or compilation errors
     * Behavior changes or incorrect wiring
     * Breaking changes to public APIs

3. **Recommended Action** - What should be done with this refactoring?
   - KEEP: Necessary + safe -> keep as-is
   - REMOVE: Unnecessary (even if safe) -> remove entirely to minimize patch
   - FIX: Necessary but unsafe -> fix the safety issues while preserving functionality

Decision Process:

For EACH detected refactoring, evaluate:
a. Locate the exact diff hunks that implement this refactoring
b. Assess necessity: Is it necessary/unnecessary for the bug fix?
c. Check safety: Are there any implementation issues?
d. If unsafe, identify specific safety issues (missing updates, type errors, etc.)
e. Recommend action based on necessity + safety combination:
   
   Necessity     Safety    -> Action
   ---------------------------------
   NECESSARY     SAFE      -> KEEP
   NECESSARY     UNSAFE    -> FIX (must preserve fix functionality)
   UNNECESSARY   SAFE      -> REMOVE (minimize patch)
   UNNECESSARY   UNSAFE    -> REMOVE (no value, has risks)

f. If action is FIX, provide specific suggestions on how to address the issues

Consistency rules:
- If you mark safety as "unsafe", action MUST be "fix" (if necessary) or "remove" (if unnecessary).
- If you mark safety as "safe", safety_issues MUST be empty.
- If action is "keep" or "remove", fix_suggestion MUST be null.
- If you cannot find the refactoring in the diff, mark it as unnecessary + unsafe and explain the mismatch in safety_issues.
- Do NOT claim a refactoring is safe while listing safety issues or requesting a fix.
- Do NOT default to "has_issues" if the evidence shows the refactoring is necessary and implemented safely.

Overall verdict rules:
- "all_safe" only if every refactoring is necessary+safe with action "keep" and no safety issues.
- "has_issues" if any refactoring is unsafe or unnecessary (action fix/remove).
- "uncertain" only if the diff is insufficient to verify refactorings or safety.

Output requirements:
- Respond ONLY with a fenced JSON block (```json ... ```).
- Include the fields:
  * "overall_verdict": "all_safe" | "has_issues" | "uncertain"
  * "confidence": "low" | "medium" | "high"
  * "refactoring_assessments": array of per-refactoring evaluations (see below)
  * "summary": brief summary of findings
  * "actions_needed": {"keep": N, "remove": N, "fix": N}

Per-refactoring assessment structure:
{
  "refactoring_type": "Extract Method",
  "location": "file.java:MethodName",
  "necessity": "necessary" | "unnecessary",
  "safety": "safe" | "unsafe",
  "safety_issues": ["issue 1", "issue 2"],  // empty if safe
  "action": "keep" | "remove" | "fix",
  "fix_suggestion": "How to fix if action is fix",  // null if not applicable
  "reasoning": "detailed explanation of assessment"
}
\end{lstlisting}

\textbf{2. Case-specific user prompt template:}
\begin{lstlisting}
You will be provided with an issue description, optional code context, the candidate patch diff, and a list of detected refactorings.

<issue>
{issue_description}
</issue>

<code>
{code_sections}
</code>

<patch>
{patch_diff}
</patch>

<detected_refactorings>
{refactoring_list}
</detected_refactorings>

Evaluation task:

1. For EACH refactoring in the list above:
   a. Locate the corresponding diff hunks (cite file names and line numbers)
   b. Assess NECESSITY: Is it necessary/unnecessary for the bug fix?
   c. Check SAFETY: Are there any implementation issues?
      - Missing call site updates?
      - Type incompatibilities?
      - Behavior changes?
   d. If unsafe, list specific safety issues
   e. Recommend ACTION based on the necessity x safety matrix:
      - Necessary + Safe -> KEEP
      - Necessary + Unsafe -> FIX (must preserve functionality)
      - Unnecessary + Safe/Unsafe -> REMOVE (minimize patch)
   f. If action is FIX, suggest how to address the issues
   g. If the refactoring cannot be located in the diff, mark it as unnecessary + unsafe and explain the mismatch

2. Provide overall assessment:
   - "overall_verdict": "all_safe" (all are safe and necessary) | "has_issues" (some need action) | "uncertain"
   - "actions_needed": Count how many refactorings need each action type

3. Be OBJECTIVE:
   - Necessary refactorings with implementation issues should be marked for FIX, not REMOVE
   - Unnecessary refactorings should be marked for REMOVE, even if they are safe
   - Focus on minimizing the patch while preserving the fix functionality
   - Keep reasoning consistent with safety/action (no contradictions)
   - If a refactoring is fully supported by the diff with no issues, mark it safe and allow "all_safe"

Respond strictly with the JSON format specified in the system prompt.
\end{lstlisting}

\subsection{Component 2: Patch Refinement}

\textbf{1. Component-specific system prompt template:}
\begin{lstlisting}
You are a senior Java engineer tasked with refining a bug-fix patch. Your goal is to apply the specified actions to each refactoring while keeping the core bug fix intact.

Key principles:

1. **Preserve the bug fix**: The primary fix must remain functional after refinement
2. **Apply actions precisely**:
   - KEEP: Leave the refactoring changes as-is
   - REMOVE: Revert only the refactoring-related changes, preserving the bug fix
   - FIX: Apply the fix_suggestion to make the refactoring safe
3. **Minimize changes**: Only modify what is necessary to apply the actions
4. **Maintain compilability**: Ensure the refined patch will compile successfully

Output requirements:
- Output ONLY a unified diff for the refined patch
- Use standard Git diff format with the following structure:
  * diff --git a/path b/path
  * (optional) index line
  * --- a/path
  * +++ b/path
  * @@ ... @@ hunks
- Each file change must include at minimum: diff --git, --- a/path, +++ b/path lines
- Wrap the entire diff in a ```diff code block
- Do NOT include explanations outside the code block
- The output must be directly applicable using "git apply" or "patch" commands
\end{lstlisting}

\textbf{2. Case-specific user prompt template:}
\begin{lstlisting}
You will be given:
1) The original patch diff
2) A list of refactoring assessments with recommended actions

<patch>
{patch_diff}
</patch>

<assessments>
{refactoring_assessments}
</assessments>

Apply the recommended actions:

For each refactoring assessment:
- If action is "keep": leave those changes unchanged
- If action is "remove": revert the refactoring-related changes while preserving the bug fix
- If action is "fix": apply the fix_suggestion to correct the safety issues

Important:
- The refined patch must still fix the original bug
- Output must be a complete Git unified diff format (with "diff --git", "index", "---", "+++" headers)
- Wrap the entire diff in ```diff ... ```
- Ensure the patch is syntactically correct and compilable
- The output must be directly applicable using "git apply" command
\end{lstlisting}

\end{document}